\documentclass{jfm}

\pdfoutput=1

\usepackage[english]{babel}
\usepackage{amssymb}
\usepackage{amsmath}
\usepackage{bm}
\usepackage{subfigure}
\usepackage{natbib}
\usepackage{graphicx}
\usepackage{blkarray}
\usepackage{array}
\usepackage{color}
\usepackage[utf8]{inputenc}
\usepackage[T1]{fontenc}

\definecolor{red}{rgb}{1,0,0}
\definecolor{green}{rgb}{0,1,0}
\definecolor{blue}{rgb}{0,0,1}
\newcommand{\red}{\color{red}}

\def\strutdepth{\dp\strutbox}
\def\nw#1{\strut\vadjust{\kern-\strutdepth\vtop to0pt{\vss\hbox to\hsize
{\hskip\hsize\hskip5pt$\leftarrow$\hss\strut}}}{\red \em #1}}

\begin{document}

\shorttitle{Rayleigh-B\'enard convection with a melting boundary} 
\shortauthor{B. Favier, J. Purseed and L. Duchemin} 

\title{Rayleigh-B\'enard convection\\ with a melting boundary}

\author{B. Favier\aff{1}\corresp{\email{favier@irphe.univ-mrs.fr}}, J. Purseed\aff{1} and L. Duchemin\aff{1}}
\affiliation{\aff{1} Aix-Marseille Universit\'e, CNRS, \'Ecole Centrale Marseille, IRPHE UMR 7342, Marseille, France}

\maketitle

\begin{abstract}
We study the evolution of a melting front between the solid and liquid phases of a pure incompressible material where fluid motions are driven by unstable temperature gradients.
In a plane layer geometry, this can be seen as classical Rayleigh-B\'enard convection where the upper solid boundary is allowed to melt due to the heat flux brought by the fluid underneath.
This free-boundary problem is studied numerically in two dimensions using a phase-field approach, classically used to study the melting and solidification of alloys, which we dynamically couple with the Navier-Stokes equations in the Boussinesq approximation.
The advantage of this approach is that it requires only moderate modifications of classical numerical methods.
We focus on the case where the solid is initially nearly isothermal, so that the evolution of the topography is related to the inhomogeneous heat flux from thermal convection, and does not depend on the conduction problem in the solid.
From a very thin stable layer of fluid, convection cells appears as the depth---and therefore the effective Rayleigh number---of the layer increases.
The continuous melting of the solid leads to dynamical transitions between different convection cell sizes and topography amplitudes.
The Nusselt number can be larger than its value for a planar upper boundary, due to the feedback of the topography on the flow, which can stabilize large-scale laminar convection cells.
\end{abstract}


\section{Introduction}

Thermally or compositionally-driven convection remains a fascinating area of research with diverse applications from geophysics, where it plays a key role in stirring the Earth's atmosphere \citep{Stevens2005} or inner core \citep{Roberts201557}, to nonlinear physics, where it is a canonical example of pattern formation and self-organization \citep{Cross1993}.
Convection is often studied in the classical Rayleigh-B\'enard (RB) configuration both due to its simplicity and well-defined control parameters \citep{Bodenschatz2000}.
Although this configuration is still actively studied and contributes to our fundamental understanding of convection processes, it is highly idealized compared to more realistic natural configurations involving non-uniform heating \citep{ROSSBY19659,killworth_manins_1980}, unsteady buoyancy forcing \citep{venezian_1969,Roppo1984,Singh2015}, complex geometries \citep{gastine_wicht_aurnou_2015,Toppa2015}, non-constant transport coefficients \citep{Tackley1996,Davaille1999}, compressible effects \citep{matthews_proctor_weiss_1995,Kogan1999,Verhoven2015}, overshooting and interactions with a stably-stratified region \citep{moore_weiss_1973,Couston2017}, etc.

Among the phenomena neglected in classical RB convection, the possibility of a non-planar boundary is particularly interesting.
The case of rough boundaries has been extensively studied due to its application to laboratory experiments \citep{du_tong_2000} while the case of large-scale topographies can significantly change the nature of convection both close to onset \citep{Kelly1978,Weiss2014} and in the super-critical regime \citep{Toppa2015,zhang_sun_bao_zhou_2018}.
While the topography is usually fixed initially, many natural mechanisms can dynamically generate non-trivial topographies.
The two-way coupling between a flow and an evolving boundary, being due to erosion, melting or dissolution, has recently received some attention \citep{claudin_duran_andreotti_2017,ristroph_2018}, and is at the origin of many geological patterns \citep{Meakinrspa20090189}.
Of interest here is the case of melting, where a natural mechanism able to dynamically generate non-trivial topographies is thermal convection itself.
It can locally melt or freeze the solid boundaries as a result of non-uniform heat fluxes.
This coupling between thermal convection and melting or freezing boundaries finds applications in various fields, from geophysics where it can affect the dynamics of the Earth's mantle and inner core \citep{Alboussiere2010,Labrosse2017}, the thermal evolution of magma oceans \citep{Ulvrova2012} or the melting of ice in oceans \citep{Martin1977,Keitzl2016}; to dendritic growth where it affects the structure of the growing solid phase \citep{BECKERMANN1999468}.

Of particular interest to the present study is the work of \cite{Vasil2011}.
They considered the gradual melting of a pure isothermal solid at the melting temperature heated from below.
As the solid melts, the liquid layer grows vertically until it reaches the critical height above which convection sets in.
The linear stability of this system is not trivial since the equilibrium background is evolving with time due to the continuous melting \citep{Walton1982}.
This has led previous authors to focus on the limit of large Stefan numbers, for which there is a time scale separation between the growth rate of the convection instability and the evolution of the background state \citep{Vasil2011}.
Many theoretical and numerical studies concerned with this problem focus on a one-way coupling where the release of latent heat affects the buoyancy of the fluid, but the dynamical effect of the topography created by this phase change is often neglected \citep{Keitzl2016}.
There exists a variety of methods to take into account the evolving phase change boundary: enthalpy methods \citep{Voller90,Ulvrova2012}, Lattice-Boltzmann approaches \citep{Jiaung2001,Babak2018}, levet set methods \citep{GIBOU2007536} and Arbitrary Lagrangian-Eulerian schemes \citep{MACKENZIE2002526,Ulvrova2012}.
Here we consider a self-consistent framework where the free-boundary problem associated with the Stefan boundary condition is solved implicitly using a phase-field method \citep{Boettinger2002}.
Adding moderate complexity to the regular Boussinesq equations, our approach is applied to the case of Rayleigh-B\'enard convection with a melting upper boundary.
We focus on the particular case where the temperature of the solid is initially close to its melting temperature.
This simple configuration does not allow for an equilibrium, since the solid phase is not cooled and will therefore continuously melt.
Note also the simultaneous and independent study by \cite{Babak2018}, who considered a similar configuration, but mostly focused on global quantities such as the heat flux or the statistical properties of the interface.
In addition to independently confirm some of their findings, we also present a detailed description of the transition between diffusive and convective regimes, we discuss the secondary bifurcation which destabilizes the initial set of convective rolls and we derive scalings for the melting velocity as a function of the Stefan number.
The case where the system is both heated from below and cooled from above which can lead to quasi-steady states, as in the experimental study of \cite{davis_muller_dietsche_1984}, will be studied later.
 
The paper is structured as follows.
The general formulation of the physical problem is presented in section~\ref{sec:model}.
We then discuss how the free-boundary conditions are treated using a phase-field method in section~\ref{sec:methods}.
The phenomenology of the melting dynamics is described in section~\ref{sec:phenomeno} and we describe quantitatively the effect of varying the Stefan number in section~\ref{sec:st}.
We finally conclude in section~\ref{sec:conclu}.

\section{Formulation of the problem\label{sec:model}}

\begin{figure}
   \vspace{5mm}
   \centering
   \includegraphics[width=0.75\textwidth]{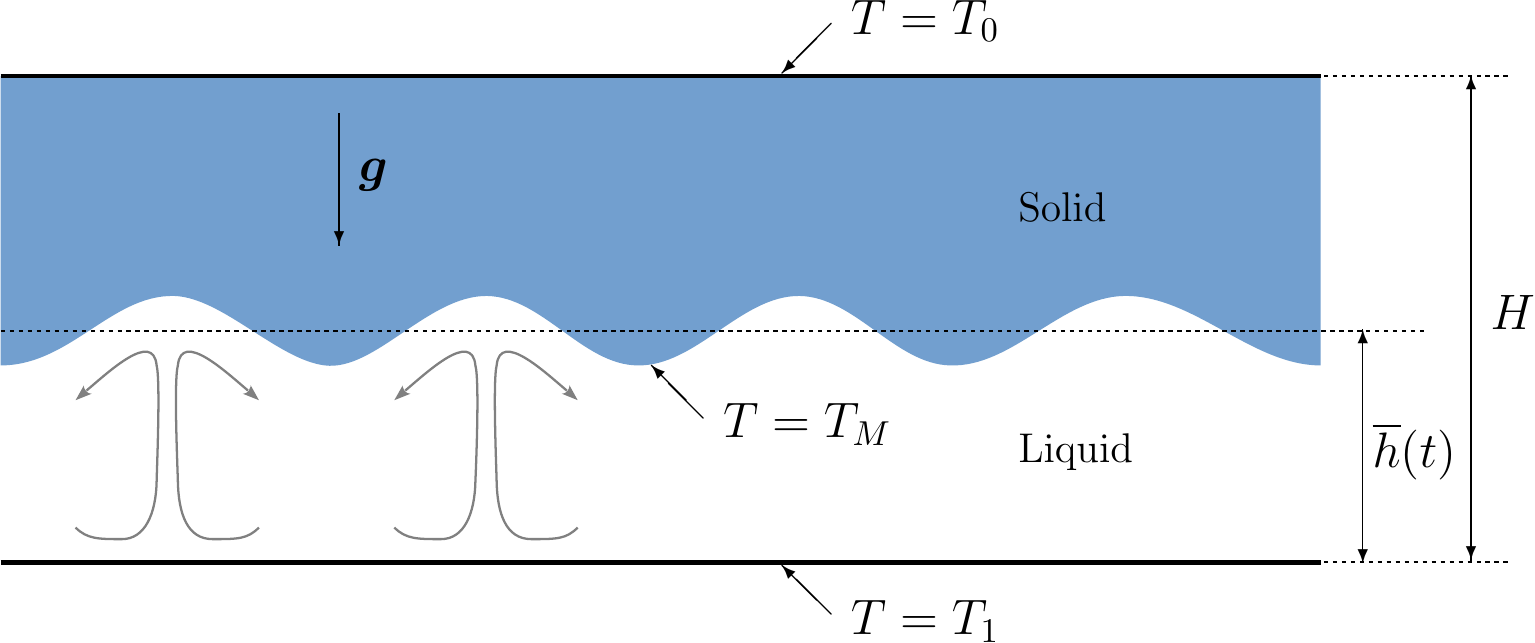}
   \caption{Schematic description of the problem considered. The blue region corresponds to the solid phase and the white region to the liquid phase. $T_M$ is the melting temperature of the pure substance. In this paper, we focus on the particular case where the solid is isothermal so that $T_M=T_0$.\label{fig:schema}}
\end{figure}

We consider the evolution of a horizontal layer of a pure incompressible substance, heated from below. The domain is bounded above and below by two impenetrable, no-slip walls, a distance $H$ apart.
The layer is two-dimensional with the $x$-axis in the horizontal direction and the $z$-axis in the  vertical direction, pointing upwards.
The gravity is pointing downwards $\bm{g}=-g\bm{e}_z$.
The horizontal size of the domain is defined by the aspect ratio $\lambda$ so that the substance occupies the domain $0<z<H$ and $0<x<\lambda H$ and we consider periodic boundary conditions in the horizontal direction.
We impose the temperature $T=T_1$ at the bottom rigid boundary and $T=T_0$ at the top rigid boundary with $T_0<T_1$.
The melting temperature $T_M$ of the substance is such that $T_0<T_M<T_1$.
Both liquid and solid phases of the substance are therefore coexisting inside the domain (see Figure~\ref{fig:schema}).
In this paper, we focus on the particular case where the solid is isothermal so that $T_M=T_0$.
For simplicity, we assume that both density $\rho$ and thermal diffusivity $\kappa_T$ are constant and equal in both phases.
The kinematic viscosity of the fluid phase $\nu$ is also assumed constant.

In the Boussinesq approximation, using the thermal diffusion time $H^2/\kappa_T$ as a reference time scale and the total depth of the layer $H$ as a reference length scale, the dimensionless equations for the fluid phase read~:
\begin{align}
\label{eq:momentum}
\frac{1}{\sigma}\left(\frac{\partial\bm{u}}{\partial t}+\bm{u}\cdot\nabla\bm{u}\right) & =-\nabla P+ Ra \; \theta \; \bm{e}_z+\nabla^2\bm{u} \\
\frac{\partial \theta}{\partial t}+\bm{u}\cdot\nabla \theta & =\nabla^2\theta \\
\label{eq:div}
\nabla\cdot\bm{u} & =0
\end{align}
where $\bm{u}=\left(u, w\right)$ is the velocity, $\theta=(T-T_0)/(T_1-T_0)$ is the dimensionless temperature and the pressure $P$ has been made dimensionless according to $P_0=\rho \kappa_T \nu / H^2$.
$Ra$ is the Rayleigh number and $\sigma$ is the Prandtl number defined in the usual way by~:
\begin{equation}
Ra=\frac{g\alpha_t\Delta TH^3}{\nu\kappa_T} \quad \textrm{and} \quad \sigma=\frac{\nu}{\kappa_T} \ .
\end{equation}
These dimensionless quantities involve $g$ the constant gravitational acceleration, $\alpha_t$ the coefficient of thermal expansion and $\Delta T=T_1-T_0$ the temperature difference between the two horizontal plates.
For numerical convenience, the Prandtl number is fixed to be unity throughout the paper.
Note that relevant applications such as the melting of ice shelves or geophysical situations involving liquid metals are respectively at high and very low Prandtl numbers.
We nevertheless choose to reduce the large parameter space by considering the standard case $Pr=1$, leaving the study of varying the Prandtl number to future works.

In the solid phase, which we assume to be non-deformable, the dimensionless heat equation simplifies to~:
\begin{equation}
\label{eq:stefan1}
\frac{\partial \theta}{\partial t}=\nabla^2\theta \ .
\end{equation}

The specificity of this configuration, compared to classical Rayleigh-B\'enard convection with a liquid phase only, lies in the boundary conditions at the interface between solid and liquid phases.
They are given by the classical Stefan conditions \citep{woods_1992,batchelor2002perspectives} which we write in dimensionless form as
\begin{align}
\label{eq:st1}
\theta & = \theta_M\\
\label{eq:st2}
St \ \bm{v}\cdot\bm{n} & =\left(\nabla \theta^{(S)}-\nabla \theta^{(L)}\right) \cdot \bm{n} \ ,
\end{align}
where $\theta_M=(T_M-T_0)/(T_1-T_0)$ is the dimensionless melting temperature ($0<\theta_M<1$), $\bm{n}$ is the local normal to the interface (pointing towards the liquid phase), $\bm{v}$ is the interface velocity and the superscript $^{(S)}$ (resp. $^{(L)}$) denotes the solid (resp. liquid) phase.
$St$ is the Stefan number and corresponds to the ratio between latent and specific heats
\begin{equation}
\label{eq:stefan}
St=\frac{\mathcal{L}}{c_p\Delta T} \ ,
\end{equation}
where $\mathcal{L}$ is the latent heat per unit mass associated with the solid-liquid transition and $c_p$ is the specific heat capacity at constant pressure of the liquid.
Since we assume there is no density variations between the solid and the liquid phases and by continuity of the normal velocity, the interface is effectively impenetrable \citep{davis_muller_dietsche_1984}.
We additionally consider the realistic case of no-slip boundary conditions on the interface.
Finally, in this general formulation, we neglect the so-called Gibbs-Thomson effects associated with the surface energy of the solid-liquid interface \citep{batchelor2002perspectives}.
Note however that the phase-field model described in the following section includes such thermodynamical effects in order to derive a continuous model of the interface dynamics.

\section{Phase-field and numerical methods\label{sec:methods}}

In this paper, we focus on a fixed-grid method \citep{Voller90} where the spatial discretization of the physical domain is fixed with time and the interface is not explicitly tracked.
Our motivation is to derive a model which can be directly implemented into any numerical code able to solve the Navier--Stokes equations in the Boussinesq approximation without major alterations.

\subsection{Phase field approach for the interface\label{sec:pf}}

In order to solve the previous dimensionless equations without having to impose the internal boundary conditions related to the interface, we introduce the continuous phase field or order parameter $\phi(x,z,t)$ such that $\phi=0$ in the solid phase and $\phi=1$ in the liquid.
A thin interface of finite width in which $\phi$ takes values between zero and unity exists between the pure solid and liquid phases.
Writing an evolution equation for the phase-field parameter can be done in several ways. The simpler derivation, which we briefly explain here, is the geometrical approach described in \cite{BECKERMANN1999468} and starts from the Gibbs--Thomson effect
\begin{equation}
\label{eq:gt1}
\frac{v_n}{\mu} = T_M - T - \frac{\sigma_s T_M}{\mathcal{L}} \kappa \ ,
\end{equation}
where $\mu$ is the mobility, $\sigma_s$ the surface tension, $\kappa$ the mean curvature of the front, and $v_n$ the normal velocity of the interface between the solid and the liquid phases. Although $\phi$ represents a finite-thickness interface, the normal velocity of the front can be related to the time-evolution of $\phi$ at a fixed value (for instance $\phi=1/2$), through the equation
\begin{equation}
\label{eq:v_n}
v_n = \frac{\partial \phi / \partial t}{\left| \nabla \phi \right|} \ .
\end{equation}
Moreover, the curvature of the front can be computed in terms of $\phi$ through
\begin{equation}
\label{eq:curvature}
\kappa = \nabla \cdot \bm{n} = 
\nabla \cdot \left( \frac{\nabla \phi}{\left| \nabla \phi \right|} \right)_{\phi=1/2} \ .
\end{equation}
Substituting equations~\eqref{eq:v_n} and \eqref{eq:curvature} into equation~\eqref{eq:gt1}, we obtain an evolution equation for $\phi$, in which the right-hand-side depends only on $\nabla \phi$ and $\nabla^2 \phi$. 
However, this equation does not have a unique stationary solution. 
Therefore, the profile for $\phi$ has to be specified, and this point is motivated by thermodynamics considerations. 

This leads us to the second approach for deriving the evolution equation for $\phi$, based on thermodynamics and described in detail in \cite{Wang1993} among others \citep{PENROSE199044,Karma1996}. 
The entropy of a given volume $V$ is represented by the functional
\begin{equation}
\label{eq:entropy}
\mathcal{S}=\int_V \left[s-\frac{\delta^2}{2}\left(\nabla\phi\right)^2\right] \textrm{d}V \ ,
\end{equation}
where $s(e,\phi)$ is the entropy density, $e$ is the internal energy density, $\phi$ the phase field and $\delta$ a constant.
The second term in the right-hand-side of equation~\eqref{eq:entropy} is analogous to the Landau-Ginzburg gradient term in the free energy and is accounting for contributions from the liquid-solid interface.
In order to ensure that the local entropy production is positive \citep{Wang1993}, the phase-field must evolve according to
\begin{equation}
\tau\frac{\partial\phi}{\partial t}= \left. \frac{\partial s}{\partial\phi} \right|_e  + \delta^2\nabla^2\phi,
\end{equation}
where $\tau$ is a positive constant.
Following the thermodynamically-consistent derivation of \cite{Wang1993}, this leads to the following dimensional phase field equation
\begin{equation}
\label{eq:dimpf}
\tau\frac{\partial\phi}{\partial t}=\delta^2\nabla^2\phi+Q(T)\frac{d p(\phi)}{d\phi}-\frac{1}{4a}\frac{d g(\phi)}{d\phi} \ ,
\end{equation}
where $Q(T)$ is defined as
\begin{equation}
\label{eq:qto}
Q(T)=\int_{T_M}^T\frac{\mathcal{L(\zeta)}}{\zeta^2} \textrm{d}\zeta \ .
\end{equation}

In the following, we assume that the latent heat $\mathcal{L}$ does not depend on temperature and that the temperature close to the interface is always approximately the melting temperature $T_M$, \textit{i.e.} $|T-T_M|\ll T_M$, so that equation~\eqref{eq:qto} can be simplified to
\begin{equation}
\label{eq:qt2}
Q(T)\approx\frac{\mathcal{L}}{T_M^2}\left(T-T_M\right) \ .
\end{equation}
Note that the validity of this simplification can be questionable in our case since thermal boundary layers will develop close to the interface.
We nevertheless checked its impact on our results by comparing the original function defined by \eqref{eq:qto} to its simplified version \eqref{eq:qt2}, and found no significant differences.


The two functions $p(\phi)$ and $g(\phi)$ must be prescribed in order to close the model.
While several choices exist in the literature, we use the prescription of \cite{Wang1993} which ensures that the solid and liquid phases correspond to $\phi=0$ and $\phi=1$, irrespective of the temperature distribution across both phases~:
\begin{equation}
g(\phi)=\phi^2\left(1-\phi\right)^2
\end{equation}
and
\begin{equation}
\label{eq:pdp}
p(\phi)=\frac{\int_0^{\phi}g(\xi)\textrm{d}\xi}{\int_0^{1}g(\xi)\textrm{d}\xi}=\phi^3\left(10-15\phi+6\phi^2\right) \ .
\end{equation}
The function $g(\phi)$ corresponds to a double-well and ensures that the phase-field is either equal to $0$ or $1$ everywhere except close to the liquid-solid interface where the phase change occurs.
The positive constant $a$ in equation~\eqref{eq:dimpf} is related to the amplitude of the potential barrier between the two equilibria.
The function $p(\phi)$ ensures a continuous transition between each extremum value of $\phi$.
Note that in a steady one-dimensional configuration, and assuming that $T=T_M$, equation~\eqref{eq:dimpf} leads to a simple analytical profile for the phase variable around the interface located at $x=x_i$ given by
\begin{equation}
\label{eq:eqp}
\phi(x)=\frac12\Big[1-\tanh\left(\frac{x-x_i}{2\sqrt{2a}\delta}\right)\Big] \ ,
\end{equation}
assuming that $\phi=1$ as $x\rightarrow-\infty$ and $\phi=0$ as $x\rightarrow+\infty$.
The diffuse interface has therefore a characteristic thickness equal to $\delta \sqrt{a}$.

The corresponding dimensional temperature equation is given by \citep{Wang1993}:
\begin{equation}
\label{eq:tempeq_dim}
\frac{\partial T}{\partial t} +\bm{u}\cdot\nabla T = \kappa_T\nabla^2 T - \frac{\mathcal{L}}{c_p}\frac{\partial p(\phi)}{\partial t} 
\end{equation}
where the last term corresponds to the release or absorption of latent heat as the phase field varies in time.
Note that the fluid is assumed to be at rest in \cite{Wang1993}, but other phase field models have since included the advection term \citep{BECKERMANN1999468,ANDERSON2000175}.
Using the same non-dimensionalization as in section \ref{sec:model}, the phase-field equation \eqref{eq:dimpf} and the temperature equation \eqref{eq:tempeq_dim} read
\begin{eqnarray}
\label{eq:pf}
\frac{\epsilon^2}{m}\frac{\partial\phi}{\partial t} & = & \epsilon^2\nabla^2\phi + \frac{\alpha\epsilon}{St} \left(\theta-\theta_M\right)\frac{dp}{d\phi} - \frac{1}{4}\frac{d g}{d\phi} \ , \\
\label{eq:tempeq_adim}
\frac{\partial\theta}{\partial t} & = &
-\bm{u}\cdot\nabla \theta + 
\nabla^2\theta-St\frac{dp}{d\phi}\frac{\partial\phi}{\partial t} \ ,
\end{eqnarray}
where
\begin{equation}
\alpha=\frac{\mathcal{L}^2 H \sqrt{a}}{\delta c_p T_M^2}
\end{equation}
is the coupling parameter between the phase field and the temperature field. 
The dimensionless interface thickness and mobility are respectively:
\begin{equation}
\epsilon=\frac{\delta \sqrt{a}}{H} \ ,
\qquad
m=\frac{\delta^2}{\tau\kappa_T} \ ,
\end{equation}
and the Stefan number is defined in equation~\eqref{eq:stefan}.
It is clear from equation~\eqref{eq:eqp} that $\epsilon$ represents the typical interface thickness in the dimensionless space.
Since we only consider cases where the bottom boundary is in the liquid phase while the top boundary is in the solid phase, we impose Dirichlet boundary conditions on the phase field
\begin{equation}
\phi\vert_{z=0}=1 \quad \text{and} \quad \phi\vert_{z=1}=0 \ ,
\end{equation}
and we recall that we impose the temperature at the boundaries
\begin{equation}
\theta\vert_{z=0}=1 \quad \text{and} \quad \theta\vert_{z=1}=0 \ .
\end{equation}

This phase-field model was initially derived in a much more general context than the classical Stefan problem, focusing on the micro-physics of solidification.
It is indeed consistent with the Gibbs--Thompson effects where the temperature at the interface is not exactly the melting temperature, but additionally depends on the local curvature and velocity of the interface.
Following the asymptotic analysis of \cite{Caginalp1989}, \cite{Wang1993} showed that in the limit of a vanishing interface thickness $\epsilon\rightarrow0$, the following boundary condition applies at the interface
\begin{equation}
\label{eq:stefan_bc}
\theta-\theta_M=-\frac{St}{\alpha}\left(\kappa+\frac{v_i}{m}\right) \ ,
\end{equation}
where the parameter $St/\alpha$ can be seen as a dimensionless capillary length, $\kappa$ is the dimensionless interfacial curvature and $v_i$ is the normal velocity of the interface.
Thus, in the additional limit where $St/\alpha\rightarrow0$, and for finite curvature and interface velocity, we recover the original Stefan boundary condition \eqref{eq:st1} where $\theta=\theta_M$ at the interface, as predicted by \cite{Caginalp1989}.
The value of the mobility is irrelevant provided that the two limits above are respected.
In conclusion, the original Stefan problem is recovered provided that
\begin{align}
\label{eq:limit1}
\epsilon & \ll 1 \\
\label{eq:limit2}
\frac{St}{\alpha} & \ll 1
\end{align}
while the mobility is fixed to be unity here.

In practice, all the additional parameters introduced by the phase-field formulation are in fact strongly constrained by the limits~\eqref{eq:limit1}-\eqref{eq:limit2}.
$\epsilon$ is typically proportional to the numerical grid size in order to accurately solve for the interface region whereas $\alpha$ is limited by stability constraints.
For the interested reader, the effect of these parameters is discussed in more details in Appendix~\ref{sec:appA}.
In the following, the interface thickness $\epsilon$ is comparable with the smallest grid size whereas $\alpha$ is typically of order $St/\epsilon$ so that both limits~\eqref{eq:limit1} and \eqref{eq:limit2} are satisfied.

\subsection{Navier--Stokes and heat equations}

The phase-field model described above satisfies the thermal Stefan conditions at the interface given by equations~\eqref{eq:st1}-\eqref{eq:st2}.
We also have to ensure that the interface corresponds to a no-slip boundary condition for the velocity.
Here we choose an immersed boundary method \citep{Mittal2005} called the volume penalization method.
The no-slip boundary condition at the liquid-solid interface is implicitly taken into account by adding a volume force to the classical Navier--Stokes equations, solved simultaneously in both liquid and solid domains, leading to
\begin{equation}
\label{eq:mompen}
\frac{1}{\sigma}\left(\frac{\partial\bm{u}}{\partial t}+\bm{u}\cdot\nabla\bm{u}\right)=-\nabla P+Ra\, \theta \, \bm{e}_z+\nabla^2\bm{u}-\frac{\left(1-\phi\right)^2\bm{u}}{\eta} \ ,
\end{equation}
where the last term is the penalization term and $\eta$ is a positive parameter.
The incompressiblity condition \eqref{eq:div} is imposed everywhere so that the total volume is necessarily conserved.
The penalized equation \eqref{eq:mompen} converges towards the Navier--Stokes equations with a no-slip boundary condition imposed at the interface \citep{Angot1999}.
The error between the original Navier--Stokes equations and their penalized version scales like $\sqrt{\eta}$ so that $\eta$ is taken as small as possible.
This ensures that this term is dominant when $\phi=0$ (\textit{i.e.} in the solid) and the velocity exponentially decays to zero on a timescale proportional to $\eta$.
When $\phi=1$ (\textit{i.e.} in the liquid), the penalization term vanishes and the regular Navier--Stokes equations \eqref{eq:momentum} are solved.
Note that the particular choice $(1-\phi)^2$ in the numerator of the penalization term is arbitrary and any continuous function that is zero in the liquid and unity in the solid is adequate.
For example, in the case of porous media, a more complex Carman--Kozeny permeability function $(1-\phi)^2/\phi^2$ can be prescribed, and the momentum and mass conservation equations can also be modified \citep{BECKERMANN1999468,LeBars2006}.
Here we choose the simplest approach of using the phase variable directly to prescribe our penalization term.
The quadratic form is chosen is order to match the Carman--Kozeny permeability for $\phi\rightarrow1$ while recovering a value of unity for $\phi\rightarrow0$.
Finally, note that all of the results discussed in this paper do not qualitatively depend on this particular prescription of the penalization term, it only affects the detailed structure of the transition between the solid and liquid phase which occurs on length scales typically smaller than the thermal and viscous boundary layers.
A comparative study of the different possible expressions for the penalization term, expressed as a function of the phase field parameter, is beyond the scope of this paper but would nevertheless prove useful to improve the convergence properties of the current model.
The penalization parameter is chosen as small as possible, noting that an explicit treatment of the penalization term leads to stability constraints, typically $\textrm{d}t<\eta$ \citep{KOLOMENSKIY20095687} where $\textrm{d}t$ is the time step.
In the following, the penalization parameter is chosen so that $\eta=2\textrm{d}t$.


We now suppose that the system is two-dimensional which naturally leads to a stream-function formulation of the Navier--Stokes equations.
The stream-function $\psi$ is defined by $\bm{u}=-\nabla\times\left(\psi\bm{e}_y\right)$ or
\begin{equation}
u=\frac{\partial\psi}{\partial z} \quad \textrm{and} \quad w=-\frac{\partial\psi}{\partial x} \ .
\end{equation}
Taking the curl of equation~\eqref{eq:momentum} and projecting onto the $y$-direction leads to the vorticity equation for a two-dimensional flow in the $(x,z)$ plane
\begin{equation}
\label{eq:sf}
\frac{\partial\nabla^2\psi}{\partial t}+\frac{\partial\psi}{\partial z}\frac{\partial\nabla^2\psi}{\partial x}-\frac{\partial\psi}{\partial x}\frac{\partial\nabla^2\psi}{\partial z}=-\sigma Ra \frac{\partial \theta}{\partial x}+\sigma\nabla^4\psi-\frac{\sigma}{\eta}\left(\nabla\times(1-\phi)^2\bm{u}\right)\cdot\bm{e}_y \ .
\end{equation}
The no-slip boundary conditions at the top and bottom boundaries correspond to
\begin{equation}
\left.\psi\right\vert_{z=0,1}=0 \quad \text{and} \quad \left.\frac{\partial\psi}{\partial z}\right\vert_{z=0,1}=0 \ .
\end{equation}
Similarly, the heat equation, including the phase-field term and the stream-function decomposition leads to
\begin{equation}
\label{eq:temp}
\frac{\partial\theta}{\partial t}+\frac{\partial\psi}{\partial z}\frac{\partial\theta}{\partial x}-\frac{\partial\psi}{\partial x}\frac{\partial\theta}{\partial z}=\nabla^2\theta-St\frac{dp}{d\phi}\frac{\partial\phi}{\partial t} \ .
\end{equation}

%
 
\subsection{Spatial and temporal discretizations}

Equations~\eqref{eq:sf}, \eqref{eq:temp} and \eqref{eq:pf} are solved using a mixed pseudo-spectral finite-difference code.
This code has been used in various context from fully-compressible convection \citep{matthews_proctor_weiss_1995,favier2012} to rapidly-rotating Boussinesq convection \citep{favier2014}.
Each variable is assumed to be periodic in the $x$-direction and is written as
\begin{equation}
f(x,z)=\sum_{n_x}\hat{f}(n_x,z)\exp\left(\textrm{i}k_xx\right)+\textrm{cc} \ ,
\end{equation}
where $n_x$ is an integer, $\textrm{cc}$ stands for conjugate terms and the wave number is defined as
\begin{equation}
k_x=\frac{2\pi n_x}{\lambda} \ .
\end{equation}
Horizontal spatial derivatives are computed in spectral space whereas vertical derivatives are discretized using a fourth-order finite-difference scheme.

For the stream-function, the dissipative fourth-order term is solved implicitly whereas the advective, temperature and penalization terms are solved explicitly.
This is achieved using a classical second-order Crank-Nicolson scheme for the implicit part coupled with a third-order Adams-Bashforth scheme for the explicit part.
For the temperature and the phase field, we use a fully explicit third-order Adams-Bashforth scheme.
An explicit treatment of these equations is indeed easier due to the nature of the coupling term on the right hand side of equation~\eqref{eq:temp}.
Note that an implicit scheme could be used to solve these equations (see for example \cite{Andersson2002}) but the stability constraint associated with solving explicitly both diffusive terms in equations~\eqref{eq:temp}-\eqref{eq:pf} is not very limiting in our two-dimensional case.
We have tested the convergence of our numerical scheme in Appendix~\ref{sec:a3}.

\section{Phenomenology of the melting dynamics\label{sec:phenomeno}}

We consider the following set of initial conditions, which only depends on the vertical coordinate $z$:
\begin{align}
\label{eq:uinit}
    \bm{u}(t=0) & = \bm{0} \\
\label{eq:tinit}
    \theta(t=0) & =   \left\{
      \begin{aligned}
        1+\left(\theta_M-1\right)z/h_0 & \quad \textrm{if} \quad z\le h_0\\
        \theta_M\left(z-1\right)/\left(h_0-1\right) & \quad \textrm{if} \quad z>h_0\\
      \end{aligned}
    \right. \\
\label{eq:pinit}
    \phi(t=0) & = \frac12\Bigg[1-\tanh\left(\frac{z-h_0}{2\sqrt{2}\epsilon}\right)\Bigg]
\end{align}
where $h_0$ is the initial position of the planar solid-liquid interface.
It corresponds to a simple piece-wise linear temperature profile with a heat flux discontinuity at $z=h_0$.
Depending on the values of $\theta_M$ and $h_0$, this can lead to situations dominated by freezing or melting.
In this paper, we focus the gradual melting of a solid that is initially nearly isothermal with a temperature close to the melting temperature.
In our dimensionless system, this corresponds to $\theta_M\ll1$.
In that configuration, no equilibrium is expected and the solid phase continuously melts until the top boundary $z=1$ is reached and only the liquid phase remains.
Note that we do not consider the limit case $\theta_M=0$ in order to avoid numerical issues in the phase field equation~\eqref{eq:pf}.
In that case, the coupling term proportional to $\theta-\theta_M$ vanishes in the whole solid which can lead to issues in the localization of the interface.
The results discussed in this paper are obtained using a typical value of $\theta_M=0.05$.
This ensures that the heat conduction in the solid plays a negligible role in the dynamics so that the evolution of the interface is solely due to the heat flux in the liquid phase (this has been checked by varying the value of $\theta_M$).

We now define several quantities that will prove useful later.
The position of the interface $h(x,t)$, which we assume to be single-valued, evolves in space and time and is implicitly defined as
\begin{equation}
    \phi(x,z=h,t)=1/2 \ .
\end{equation}
It is useful to define the effective Rayleigh number of the fluid layer, based on the actual temperature gradient across the depth of the fluid layer
\begin{equation}
\label{eq:era}
    Ra_e=Ra\left(1-\theta_M\right)\overline{h}^3 \ ,
\end{equation}
where we introduce the averaged fluid height defined as
\begin{equation}
    \overline{h}(t)=\frac{1}{\lambda}\int_0^{\lambda}h(x,t)\textrm{d}x \ ,
    \label{meanheight}
\end{equation}
where $\lambda$ is the dimensionless horizontal length of the domain.
In the following, the operator $\overline{\ \cdot \ }$ corresponds to a horizontal spatial average.
For simplicity, and by analogy with classical Rayleigh-B\'enard convection, we only work with the heat flux injected at the bottom boundary.
The heat flux consumed at the solid-liquid interface to melt the solid could equally be used, although it is more complicated to measure  numerically.
A detailed discussion of the different measures of the heat flux in this system can be found in \cite{Babak2018}.
The heat flux injected into the fluid is
\begin{equation}
\label{eq:thf}
    Q_W=-\frac{1}{\lambda}\int_{0}^{\lambda} \frac{\partial\theta}{\partial z} \bigg\rvert_{z=0} \textrm{d}x
\end{equation}
so that the Nusselt number can be defined in first approximation (see section~\ref{sec:hflux} for a more detailed discussion) by
\begin{equation}
\label{eq:nuss}
    Nu=\frac{Q_W}{Q_D}=\frac{Q_W\overline{h}}{1-\theta_M}
\end{equation}
where we have introduced the reference diffusive heat flux $Q_D$ which can be approximated for now by $(1-\theta_M)/\overline{h}$.

\begin{table}
 \begin{center}
  \begin{tabular}{ccccccccc}
    Case  & $N_x$ & $N_z$ & $Ra$ & $St$ & $\theta_M$ & $\lambda$ & $\epsilon$ & $\alpha$ \\[3pt]
      A & $\quad512\quad$ & $\quad256\quad$ & $\quad15180\quad$ & $10$ & $\quad0.1\quad$ & $\quad8\quad$ & $\quad4\times10^{-3}\quad$ & $2500$\\
      B & $256$ & $256$ & $6\times10^5$ & $10$ & $0.05$ & $1$ & $2\times10^{-3}$ & $5000$\\
      C & $4096$ & $1024$ & $10^8$ & $1$ & $0.05$ & $8$ & $10^{-3}$ & $1000$\\
      D & $1024$ & $512$ & $10^7$ & $[0.02:50]$ & $0.05$ & $6$ & $3\times10^{-3}$ & $[10:20000]$\\
  \end{tabular}
  \caption{List of numerical parameters for the different cases discussed in this study.\label{tab:one}}
 \end{center}
\end{table}

\subsection{Critical Rayleigh number\label{sec:crit}}

We focus here on the transition between a purely diffusive regime and a convection regime as the fluid depth increases with time.
We therefore consider the case where the initial height $h_0$ is small enough so that the initial conditions given by equations~\eqref{eq:uinit}-\eqref{eq:pinit} are stable.
It has been showed by \cite{Vasil2011} that the convection threshold can be modified compared to the classical Rayleigh-B\'enard problem and that a morphological mode grows as soon as $Ra(1-\theta_M)\overline{h}^3>Ra_c\approx1295.78$.
This corresponds to a significant modification of the stability criterion compared to the case of classical no-slip RB convection, for which $Ra_c\approx1707.76$ \citep{Chandra1961}.
The most unstable wave number is also reduced from $k_c\approx3.116$ to $k_c\approx2.552$.
These results are however only valid in the asymptotic limit of large Stefan numbers.
While this regime is virtually impossible to reach numerically using the current approach (there exists a time-scale separation between the dynamics of the flow and that of the interface), we nevertheless explore this critical transition for finite Stefan numbers in the following.

\begin{figure}
   \vspace{5mm}
   \centering
   \includegraphics[width=0.8\textwidth]{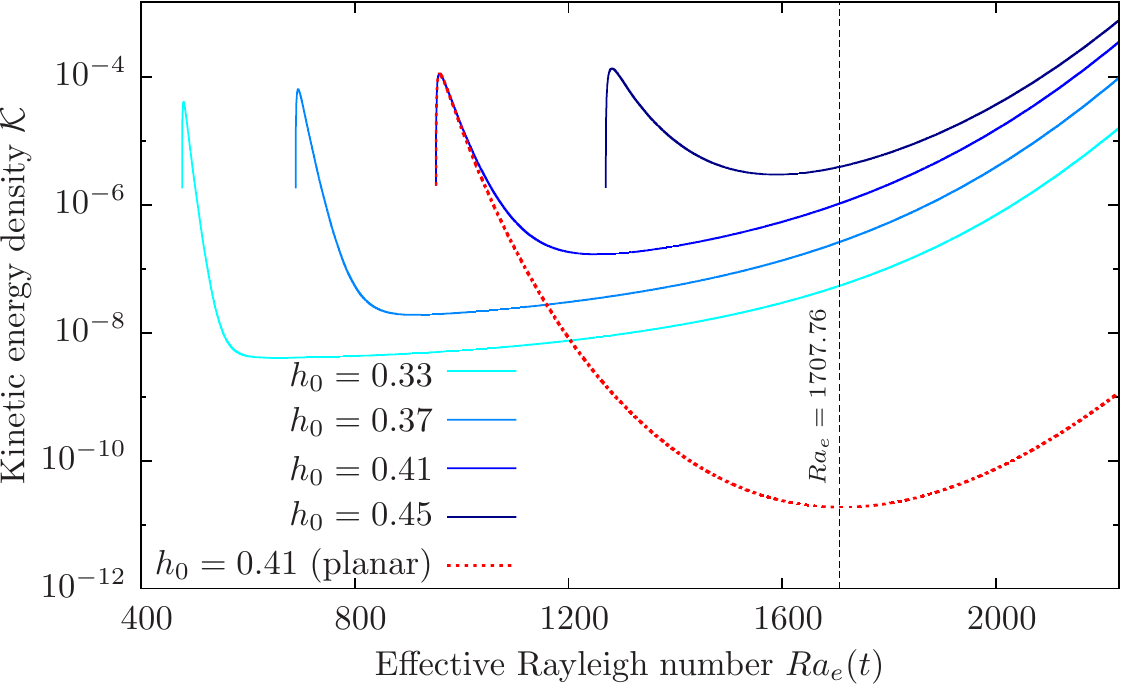}
   \caption{Kinetic energy density in the fluid domain versus the time varying Rayleigh number defined by equation~\eqref{eq:era}. Each curve corresponds to a different initial fluid depth $h_0$ in equations~\eqref{eq:tinit}-\eqref{eq:pinit}. The dashed curve corresponds to the case where only the horizontally-averaged component of the phase field is solved (\textit{i.e.} the upper boundary is effectively planar).\label{fig:threshold}}
\end{figure}

We start from the initial conditions defined by equations~\eqref{eq:uinit}-\eqref{eq:pinit} with various initial heights from $h_0=0.33$ to $h_0=0.45$, and $\theta_M=0.1$.
Using a global Rayleigh number of $Ra=15180$, this leads to an initial effective Rayleigh number varying between $Ra_e(t=0)=491$ and $1245$.
We start from infinitesimal temperature perturbations in the liquid layer only.
We consider a case where $St=10$ which is large enough to get a reasonable timescale separation while still being accessible numerically.
The other numerical parameters are given in Table~\ref{tab:one} and correspond to case A.
We define the kinetic energy density in the system by
\begin{equation}
    \mathcal{K}(t)=\frac{1}{V_f}\int_{V_f}\bm{u}^2 \textrm{d}V \ ,
\end{equation}
where $V_f(t)$ is the volume of fluid as a function of time.
The time evolution of the kinetic energy density versus time for various initial heights $h_0$ is shown in Figure~\ref{fig:threshold}.
Initially, the kinetic energy in the system briefly increases.
This is a consequence of our choice of initial conditions for which the fluid is at rest and temperature perturbations only are added.
After this short transient, the kinetic energy density decreases with time for all cases.
Surprisingly, the kinetic energy starts to grow for different effective Rayleigh numbers in each case, as early as $Ra_e\approx650$ for the smallest initial height of $h_0=0.33$.
The growth rate of this first phase is however much weaker that the typical growth rate when the effective Rayleigh number becomes larger that the classical value of $1707.76$.
We therefore do not observe a clear transition between stable and unstable behaviours at a given critical Rayleigh number, which seems to indicate that perturbations can grow at any value of the effective Rayleigh number.

In order to show that this is a consequence of the upper boundary not being exactly planar, we perform an additional simulation with the exact same parameters as above and for $h_0=0.41$.
The only difference is that we artificially smooth the upper boundary by only solving the horizontally-averaged value of the phase field (this is performed in Fourier space by truncating all modes except $k_x=0$).
This is of course artificial but nevertheless useful to understand the origin of this early growth.
The time evolution of the kinetic energy density is shown in Figure~\ref{fig:threshold} as a dashed curve.
At early times, there are no noticeable differences between the regular simulation and the artificial planar case.
However, at later times, there is no growth for the planar case until the critical value of $Ra_e\approx1710$ is reached.
This clearly shows that the very early growth of the kinetic energy is associated with the presence of a topography.
This topography is very small in amplitude since it is generated by the initial perturbations, but is nevertheless measurable numerically.
It is known that any non-planar topography will drive a baroclinic flow at any Rayleigh number \citep{Kelly1978}.
The amplitude of this gravity current scales linearly with the Rayleigh number and linearly with the amplitude of the topography and is directly forced by the misalignment between the hydrostatic pressure gradient and the inclined temperature gradient normal to the boundary.
As we get closer to the threshold $Ra_e=1707.76$, we eventually recover the classical instability mechanism of convection through an imperfect bifurcation \citep{Coullet1986}.
This is consistent with the slow growth of kinetic energy we observed in Figure~\ref{fig:threshold}, followed by an exponential phase (the growth rate for $Ra_e>1707.76$ is actually super-exponential since the Rayleigh number keeps increasing while the instability develops).

There are several reasons why we do not recover the result of \cite{Vasil2011} who found a critical transition at $Ra_e\approx1295$.
First, we are not in the asymptotic regime of large Stefan numbers.
We repeated the previous simulations at higher $St$, up to $St=100$, without qualitative changes in the results discussed above.
It is however possible that the asymptotic regime discussed by \cite{Vasil2011} is only achieved at much higher Stefan numbers.
In addition, \cite{Vasil2011} used slightly different boundary conditions and they focused on modes growing on the very slow melting timescale, which is difficult to isolate in our finite Stefan number simulations.
Finally, even if we varied all the numerical parameters of our model to confirm that the results discussed above are numerically converged, we cannot discard the possibility that the phase-field approach is inappropriate to study the evolution of infinitesimal perturbations of the topography, as it is the case here.
One must remember that the interface is here continuous with a typical width that is here much larger that the perturbations responsible for driving the baroclinic flow.
We however note that once the classical convection instability sets in, all the previous simulations starting from various initial heights lead to the same nonlinear state (apart from a temporal shift as seen in Figure~\ref{fig:threshold} at late times) which is discussed in the following sections.

\subsection{Nonlinear saturation close to onset and secondary bifurcation\label{sec:bif}}

We now explore the evolution of the system once the initial instability saturates and leads to a steady set of convective rolls.
In the following, we consider a particular case with a relatively large Stefan number $St=10$ so that we get a reasonable timescale separation between the flow and the interface dynamics.
This particular choice is made to simplify the analysis of the bifurcation to be discussed below.
For the same reason, we consider a laterally confined case where $\lambda=1$.
The other parameters are given in Table~\ref{tab:one} for case B.
We start from an initial height of $h_0=0.13$ so that the initial fluid layer is stable ($Ra_e(h_0)\approx1245$).

\begin{figure}
   \vspace{5mm}
   \centering
   \includegraphics[width=0.44\textwidth]{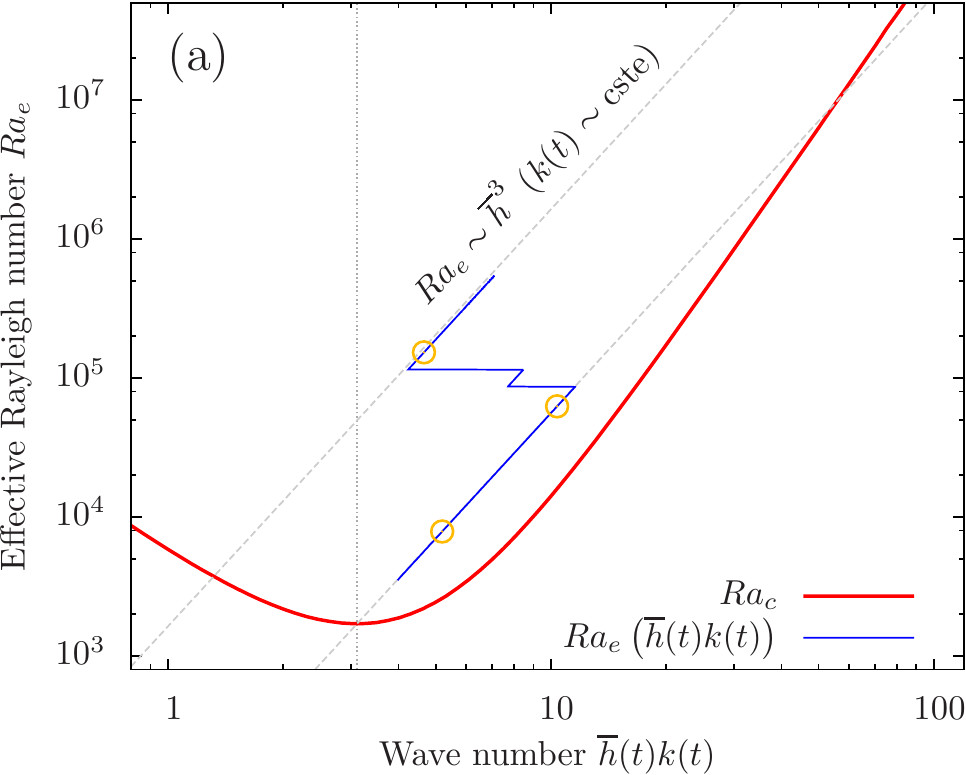}
   \hfill
   \includegraphics[width=0.53\textwidth]{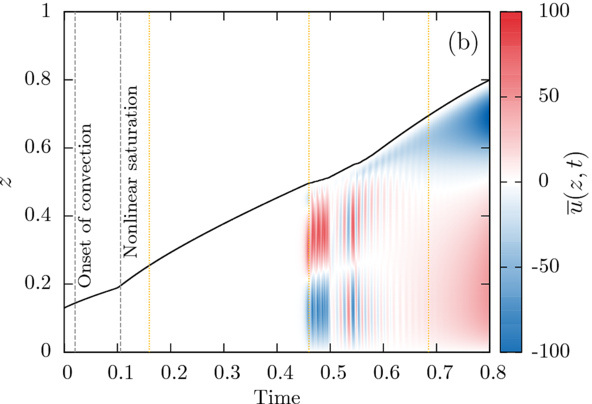}\\
   \vspace{3mm}
   \includegraphics[width=0.28\textwidth]{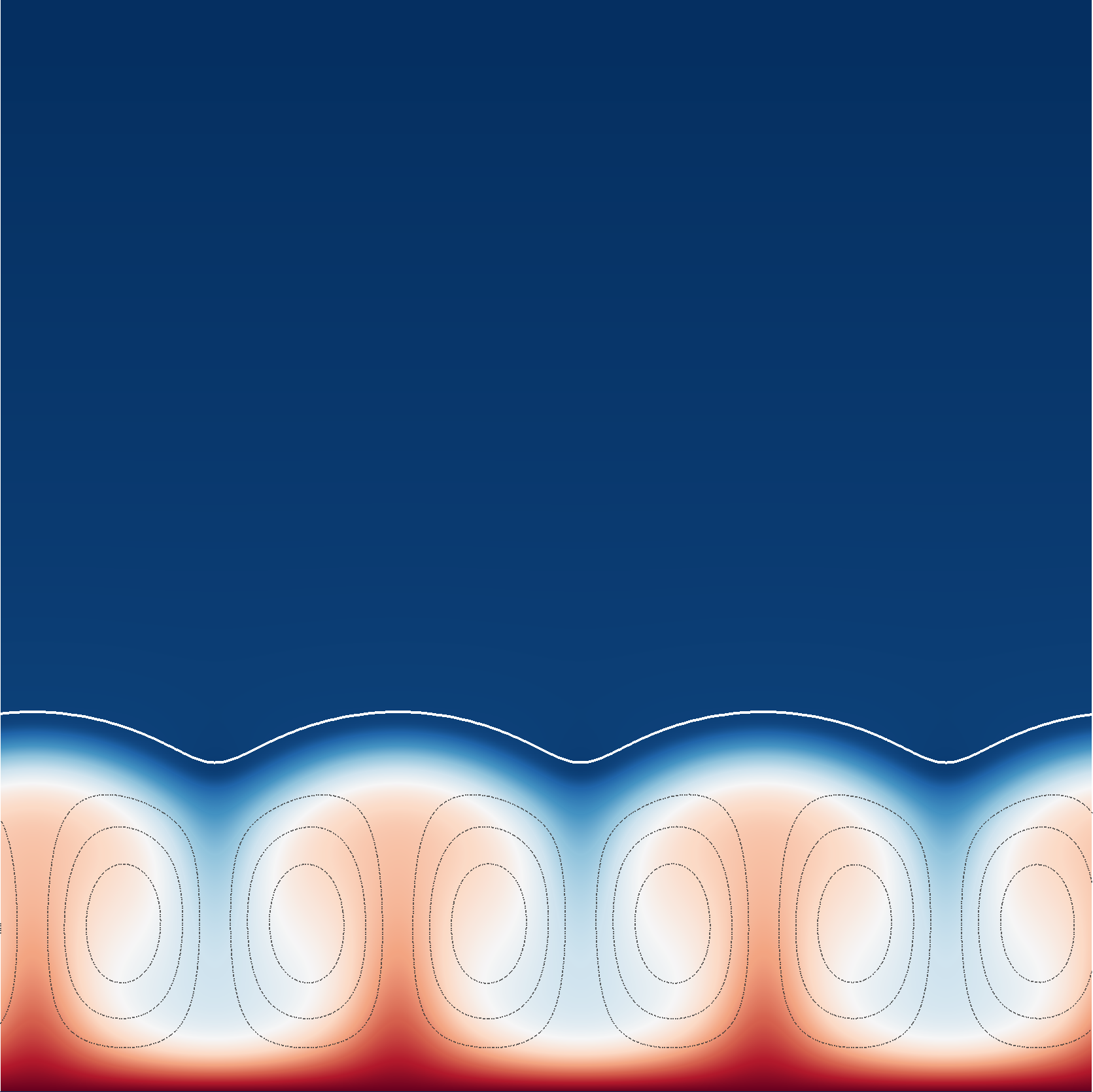}
   \hfill
   \includegraphics[width=0.28\textwidth]{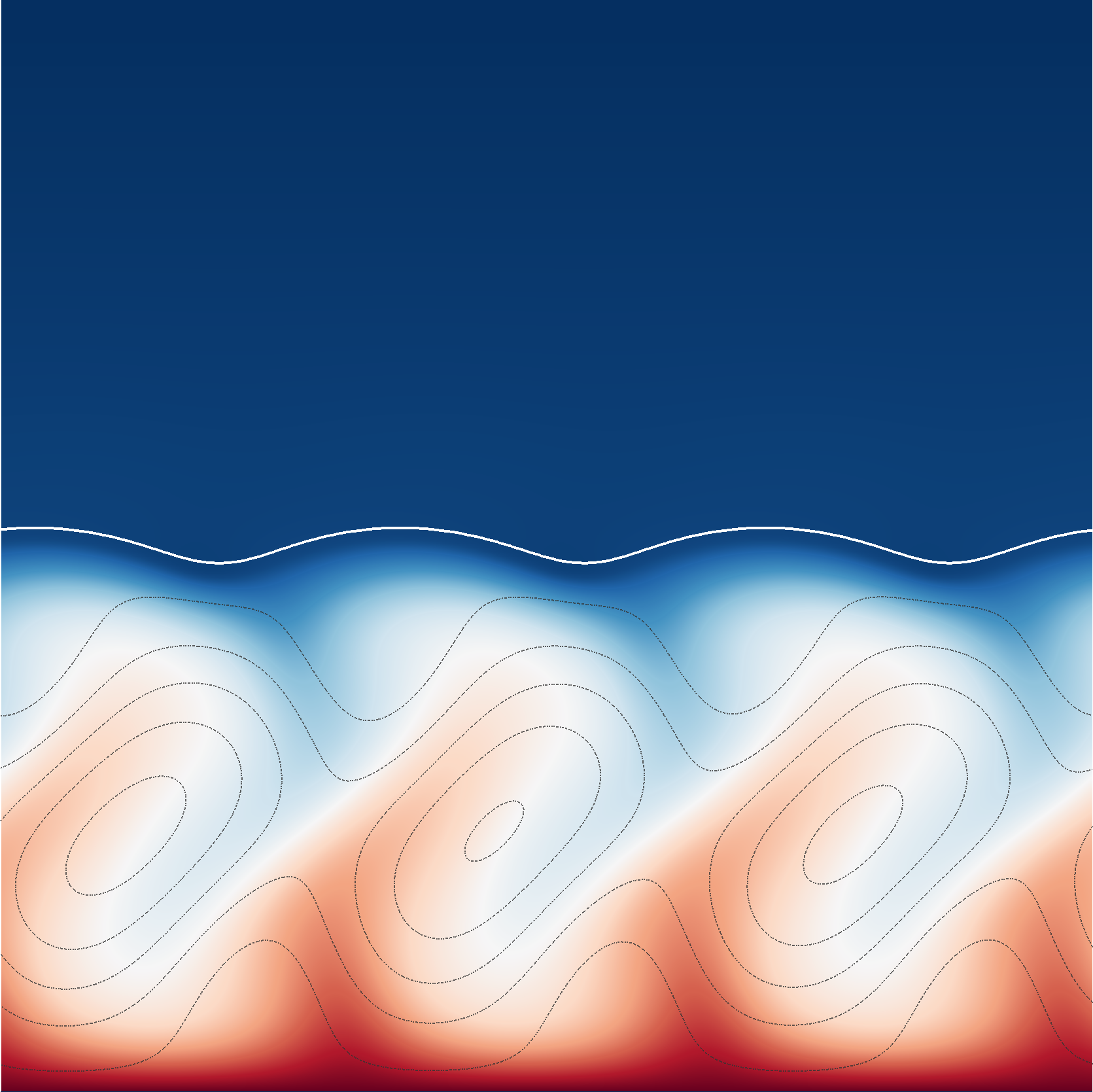}
   \hfill
   \includegraphics[width=0.28\textwidth]{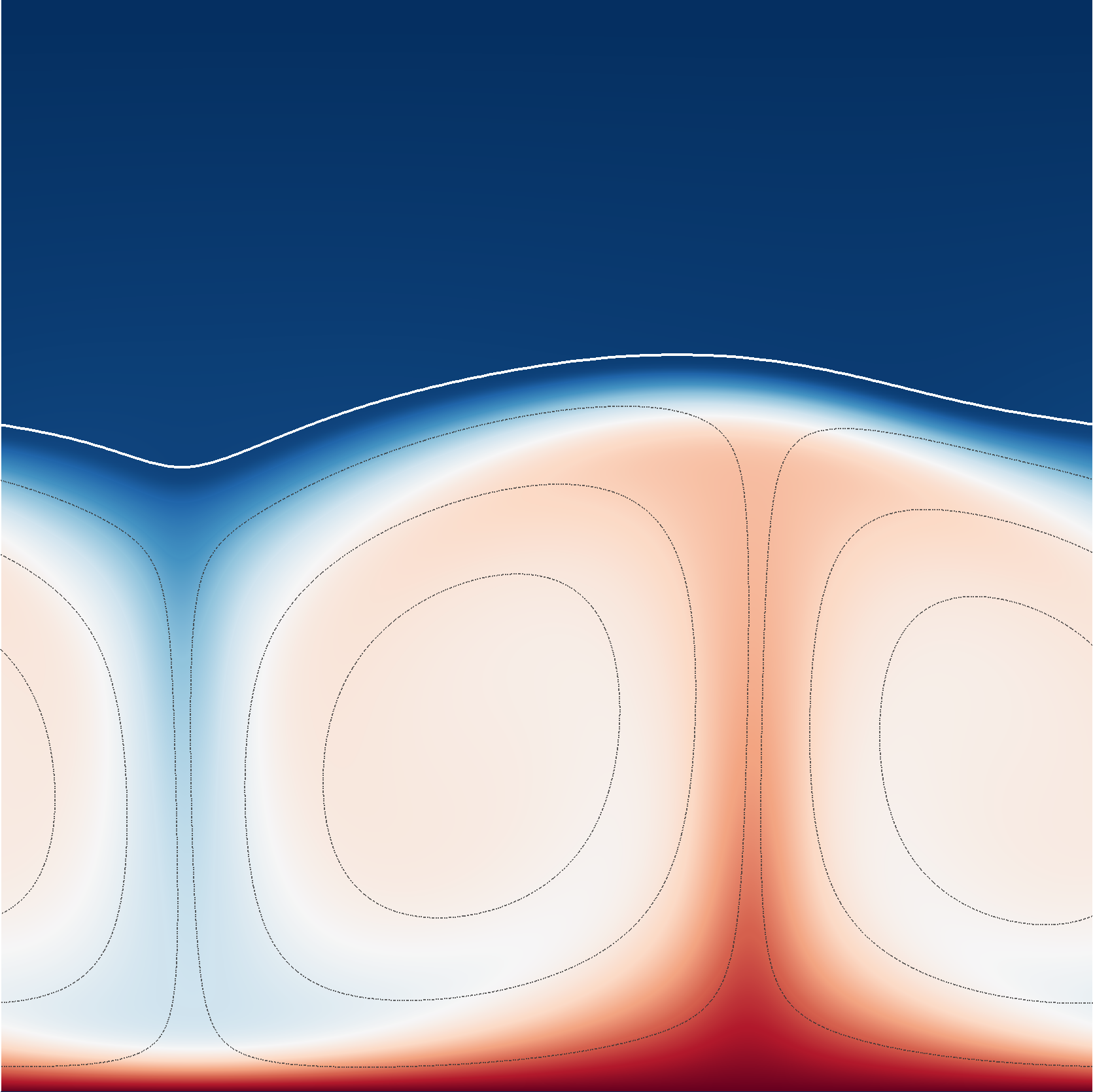}
   \caption{Generation of mean horizontal shear flows during the collapse of steady convective rolls. (a) Effective Rayleigh number as a function of the normalized wave number $\overline{h}k$. The red curve corresponds to the marginal curve for classical RB \citep{Chandra1961}. The oblique dashed lines correspond to a constant horizontal wave number and follow $Ra_e\sim \overline{h}^3$. (b) Horizontally-averaged flow $\overline{u}$ versus $z$ and time. The black line corresponds to the maximum height $\textrm{max}(h(x,t))$. The onset of convection and the nonlinear saturation are indicated with vertical dashed lines. Bottom row: temperature field shown at three successive instants, shown as empty symbols and vertical dotted lines in (a) and (b). The white line corresponds to the interface defined as $\phi=1/2$ and the grey lines correspond to streamlines.\label{fig:bifurc}
   }
\end{figure}

After the transient growth discussed in the previous section, we observe at saturation a steady flow and a significant topography which is now clearly non-planar (see Figure~\ref{fig:bifurc}).
Perhaps unsurprisingly, the wavelength of this topography is equal to that of the convective rolls below.
The solid is locally melting just above rising hot plumes but less so above sinking cold plumes.
Once this nonlinearly equilibrated set of rolls and their associated topography exist, the horizontal wavenumber of the rolls is fixed while the average fluid depth keeps increasing with time.
This can be seen by measuring the typical horizontal wavelength of the topography as the distance between two local minima of $h(x,t)$.
In order to compare with classical RB convection, we normalize the corresponding wavenumber $k$ by the time dependent averaged fluid depth $\overline{h}(t)$.
We show in Figure~\ref{fig:bifurc}(a) the effective Rayleigh number of the fluid layer as a function of this normalized wavenumber $\overline{h}k$.
The marginal stability curve of classical RB with no-slip and fixed temperature walls is shown for reference \citep{Chandra1961}.
Since the average fluid depth $\overline{h}$ increases, while the horizontal wave number of the convection remains constant, the effective Rayleigh number continuously increases like $Ra_e\sim\overline{h}^3$.
Our simulation closely follows this prediction, as shown in Figure~\ref{fig:bifurc}(a).

A simple question now arises: how long can this dynamically evolving set of convective rolls persists against the continuous vertical stretching of the fluid domain?
One possibility would be to assume that the rolls are vertically elongated until they become stable again.
This is indeed possible since the marginal curve behaves like $Ra_c\sim(\overline{h}k)^4$ for large wave numbers whereas the rolls with fixed horizontal wave number follow a $Ra_e\sim(\overline{h}k)^3$ scaling, so that they will eventually become stable as shown in Figure~\ref{fig:bifurc}(a).
This is not what is observed in the simulation however, and a bifurcation occurs well before the possible restabilisation of the initially unstable mode.
This bifurcation occurs after the rolls have been elongated vertically by an approximate factor of $3$ and corresponds to an abrupt reduction in the horizontal wavenumber $k(t)$ of the convection rolls.

The detailed nature of this bifurcation can be qualitatively understood by following the time evolution of the horizontally-averaged mean flow $\overline{u}(z,t)$, showed in Figure~\ref{fig:bifurc}(b).
This mean flow remains negligible at early times but abruptly grows when the rolls become elongated enough.
It first appears as a shear flow with one vertical wave length, effectively shearing the first set of rolls, as can be seen in the temperature field showed in the bottom middle panel of Figure~\ref{fig:bifurc}.
Once the initial set of rolls has been disrupted, the mean flow undergoes damped oscillations.
A new set of rolls is then generated by the convection instability, with a larger horizontal wavelength, therefore maintaining the unit aspect ratio of the convective cells.
This bifurcation is also visible in Figure~\ref{fig:bifurc}(a) where a jump between two $k\sim\textrm{cste}$ curves is observed.
This is reminiscent of the generation of mean shear flows in laterally confined classical RB convection \citep{Busse1983,Prat1995,goluskin_johnston_flierl_spiegel_2014,Fitzgerald2014}.
Our case is however slightly more complicated since the volume of fluid increases with time and the upper topography is non-planar.
The generation of mean horizontal shear flows is nevertheless a generic mechanism in RB convection that we also observe in our particular system.
Note that the detailed properties of this bifurcation depends on the aspect ratio of numerical domain $\lambda$, the Stefan number $St$ and the nature of the initial perturbations.

This transition between convection rolls of different sizes is expected to repeat itself as the fluid depth keeps increasing, with the additional complication that the effective Rayleigh is ever increasing due to the gradual melting of the solid, so that the bifurcation is expected to become more and more complicated.

\subsection{Behaviour at large $Ra$}

We now focus on a representative example at high Rayleigh number for which we fix $St=1$, $Ra=10^8$ and $\sigma=1$.
This simulation corresponds to case C in Table~\ref{tab:one}.
We consider a domain with a large aspect ratio of $\lambda=8$.
Doing so, we aim at minimizing the horizontal confinement effect associated with our periodic boundary conditions.
Note that this is the global aspect ratio including the solid domain, the actual aspect ratio of the liquid domain is initially much larger.
We therefore expect the liquid phase to display spatio-temporal chaos instead of purely temporal chaos typical of laterally-confined systems \citep{Manneville2006}.
The initial position of the interface is $h_0\approx0.02$ so that the effective Rayleigh number of the fluid layer is $Ra_e(t=0)=Ra(1-\theta_M)h_0^3\approx760$, well below the critical value.
Using a horizontal resolution of $N_x=4096$ and a vertical resolution of $N_z=1024$, the smallest grid size is typically $\text{d}x\approx2\times10^{-3}$ so that interface width is fixed to $\epsilon=10^{-3}$.
Assuming a Prandtl-Blasius scaling \citep{grossmann_lohse_2000}, the dimensionless width of the viscous boundary layers $\delta_v$ scales like $Ra^{-1/3}$ which typically leads to $\delta_v\ge2.2\times10^{-3}$ in our case.
Thus, the interface width $\epsilon$ is always significantly smaller that the viscous boundary layer $\delta_v$.

\begin{figure}
   \vspace{4mm}
   \centering
   \includegraphics[width=1.0\textwidth]{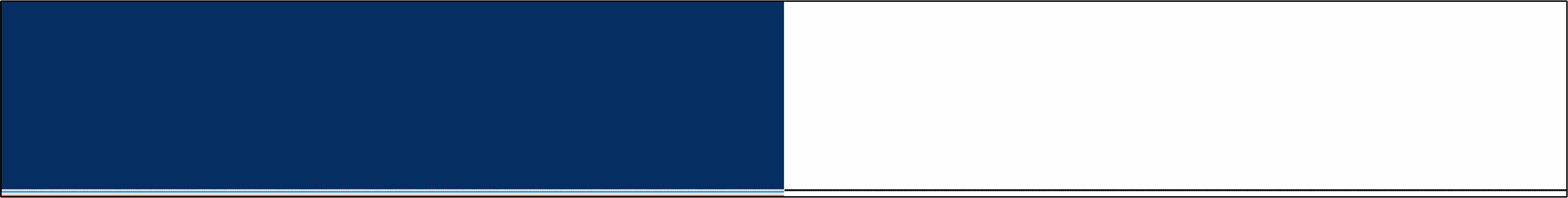}\\
   \vspace{1mm}
   \includegraphics[width=1.0\textwidth]{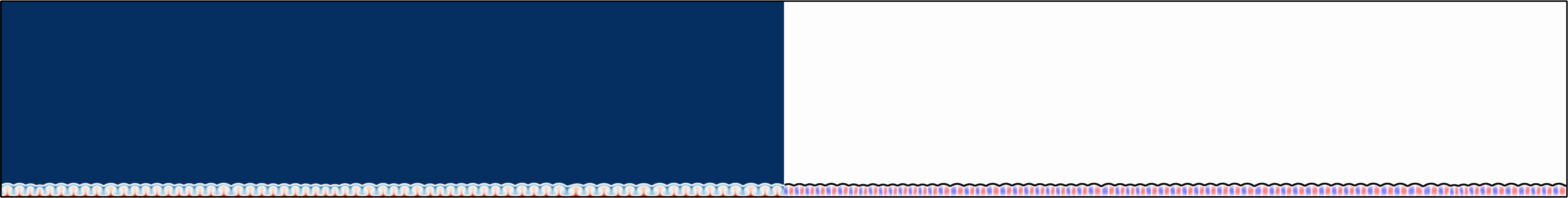}\\
   \vspace{1mm}
   \includegraphics[width=1.0\textwidth]{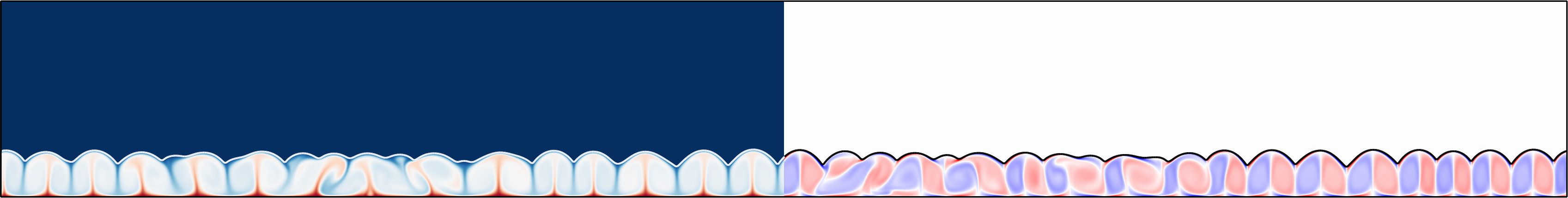}\\
   \vspace{1mm}
   \includegraphics[width=1.0\textwidth]{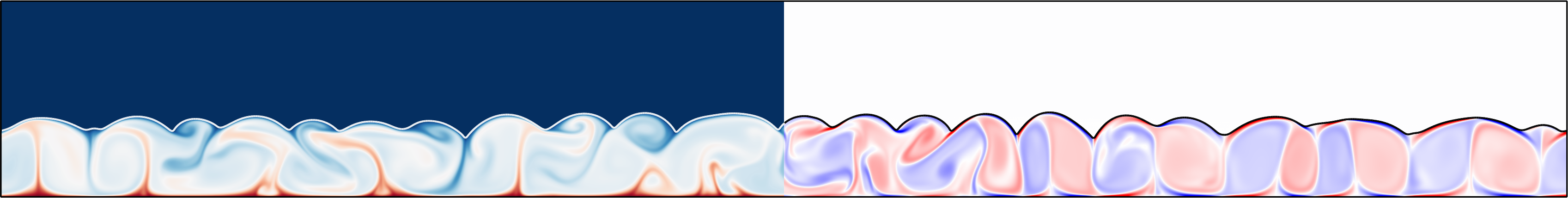}\\
   \vspace{1mm}
   \includegraphics[width=1.0\textwidth]{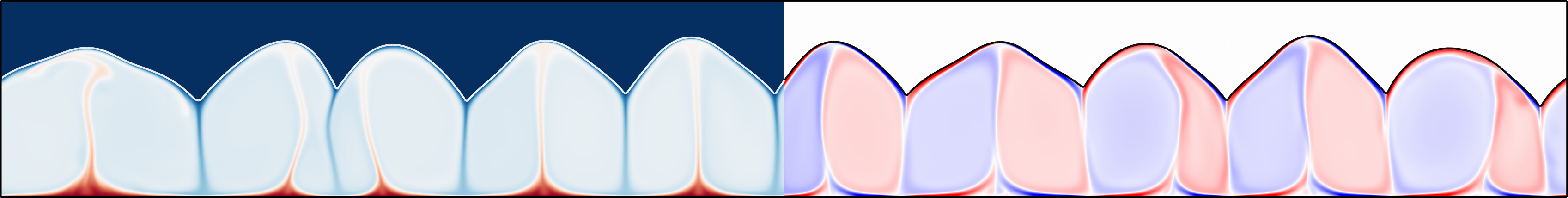}\\
   \vspace{1mm}
   \includegraphics[width=1.0\textwidth]{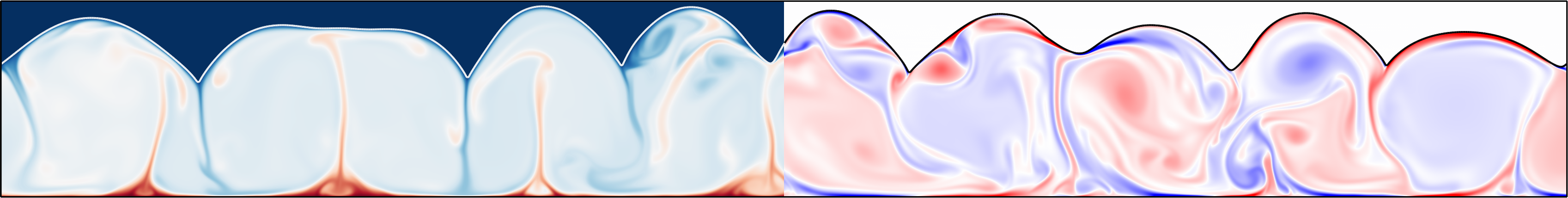}
   \caption{\label{fig:visus} Visualizations of the total numerical domain for case C in Table~\ref{tab:one}. The temperature is shown on the left (dark red corresponds to $\theta=1$ while dark blue corresponds to $\theta=\theta_M$) while vorticity is shown on the right (blue and red colors correspond to $\pm 0.25 \ \omega_{\text{max}}$ respectively). The grey line corresponds to the interface defined by the isosurface $\phi=1/2$. Time is increasing from top to bottom: $t=5\times10^{-4}$, $1.5\times10^{-3}$, $6\times10^{-3}$, $1.2\times10^{-2}$, $2.4\times10^{-2}$ and $3\times10^{-2}$. See also Movie1 in Supplementary materials.}
\end{figure}

The simulation is run until the interface reaches the upper boundary $z=1$.
For the parameters of this simulation, this approximately takes $0.03$ thermal diffusive timescales.
We first show in Figure~\ref{fig:visus} (see also Movie1 in Supplementary materials) visualizations showing the temperature and vorticity distributions at different times during the simulation.
At the early times, the solution is purely diffusive until the liquid depth reaches its critical value above which convection sets in.
Convection is initially steady and laminar, as observed previously, with approximately $132$ convective cells across the whole domain.
As the interface progresses, this initial set of convective rolls is vertically stretched, eventually forcing a secondary transition leading to larger convective cells, as discussed previously in section~\ref{sec:bif}, although the nature of the bifurcation appears to be different in this large aspect ratio domain (see below).
This alternation between quasi-stationary phases of melting where the number of convective cells is conserved and more violent transitions associated with a reordering of the convective cells continues until the upper boundary is eventually reached.

\begin{figure}
   \vspace{5mm}
   \centering
   \includegraphics[width=0.9\textwidth]{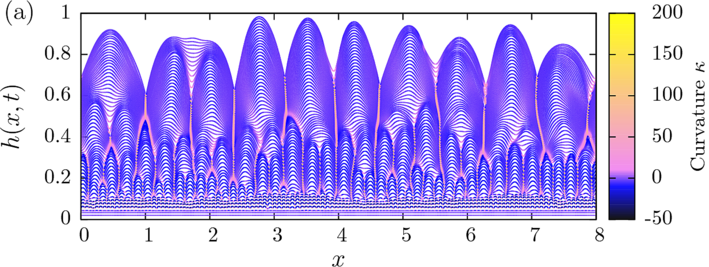}
   \includegraphics[width=0.95\textwidth]{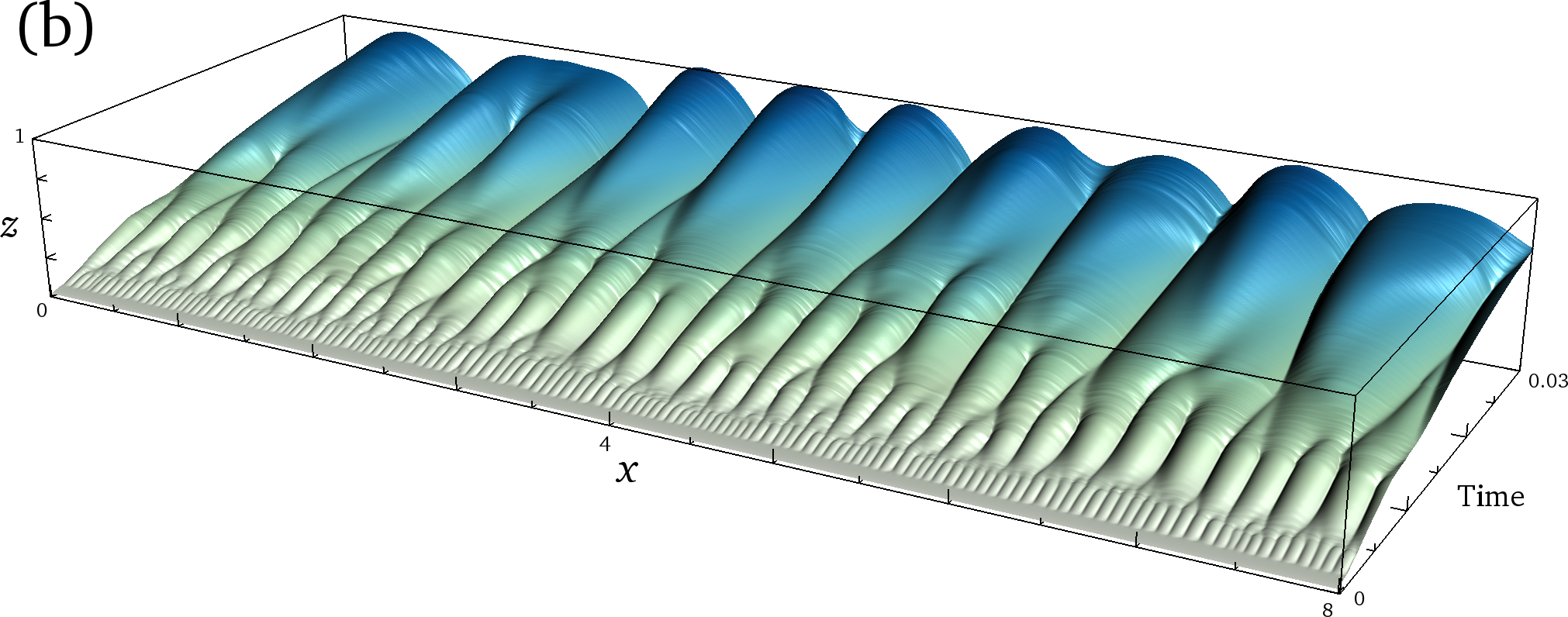}
   \caption{(a) Position of the interface $h(x,t)$ as a function of $x$ for different times separated by approximately $5\times10^{-4}$ thermal diffusion times. The color of the curves correspond to the signed curvature defined by equation~\eqref{eq:curv}. Dark colors correspond to small negative curvatures whereas light colors correspond to cusps with large positive curvatures. (b) Three-dimensional view of the spatio-temporal evolution of the interface $h(x,t)$. The color corresponds to the local value of $h(x,t)$.\label{fig:htime}}
\end{figure}

Let us first discuss the shape of the interface as the solid continuously melts.
We first show in Figure~\ref{fig:htime}(a) the interface position as a function of the horizontal coordinate $x$ at different times.
The interface is obtained by interpolating the phase field variable in order to find the iso-contour $\phi=1/2$.
Equivalently, the interface can be defined as the isotherm $\theta=\theta_M$, which leads to the same results, apart from very localized regions of high curvatures where a slight mismatch between the two isocontours is observed, as expected from the Gibbs-Thomson relation~\eqref{eq:stefan_bc}.
These discrepancies are however negligible here (\textit{i.e.} the maximum distance between the isocontours $\phi=1/2$ and $\theta=\theta_M$ is smaller than the thickness of the boundary layers), as expected from our choice of large coupling parameter $\alpha$.
The color of the curves in Figure~\ref{fig:htime}(a) corresponds to the signed curvature, derived from the interface position $h(x,t)$ following
\begin{equation}
\label{eq:curv}
    \kappa(x,t)=\frac{\partial_{xx}h}{\left[1+\left(\partial_xh\right)^2\right]^{3/2}} \ .
\end{equation}
The maximum value of the curvature corresponds to cusps joining two cavities of the topography and is approximately $\kappa_{\text{max}}\approx300$.
This is of the same order than the largest curvature achievable by our phase-field approach, which can be approximated by the inverse of the interface width $\epsilon^{-1}=10^3$.
We can therefore be confident that the cusps are numerically resolved and not artificially smoothed by our diffuse interface approach.
The horizontal positions of these cusps appear to be very stable, which corresponds to a spatial locking between the convection rolls and the topography \citep{Vasil2011}.
One can also note that these cusps often correspond to small melting rates (\textit{i.e.} the successive profiles of $h(x,t)$ are close) compared to the much larger cavities with negative curvature where intense localized melting events driven by the underlying hot thermal plumes are observed.
An alternative three-dimensional view of the spatio-temporal evolution of the interface is also shown in Figure~\ref{fig:htime}(b).
The successive bifurcations between different rolls size is clearly visible.
We observe various types of cell merging events, from two adjacent cells merging into one, to more complicated behaviours where one cell disappears leading to the merging of its neighbour cells.

\begin{figure}
   \vspace{5mm}
   \centering
   \includegraphics[width=0.98\textwidth]{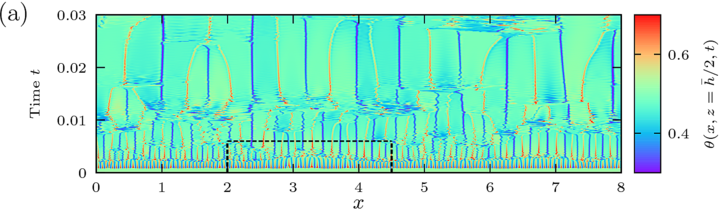}\\
   \includegraphics[width=0.99\textwidth]{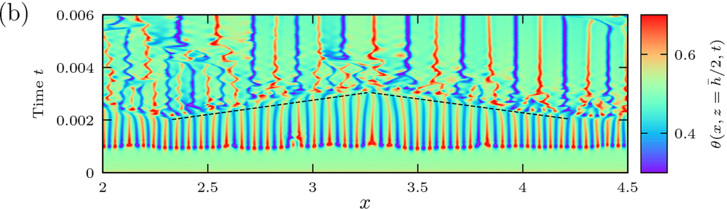}
   \caption{Spatio-temporal evolution of the temperature at the middle of the fluid layer $\theta(x,\overline{h}/2,t)$. (a) Full duration of the simulation $0<t<3\times10^{-2}$ and full spatial extent $0<x<8$. The dashed line corresponds to the zoom shown below. (b) Zoom in at early times $0<t<6\times10^{-3}$, the dashed lines follow the propagation of a defect in the initially-steady array of convective rolls.\label{fig:tmid}}
\end{figure}

We now describe the dynamics of the fluid flow, which is strongly correlated with that of the interface.
Our system is not laterally confined so that there is no significant horizontal mean flow, as seen in section~\ref{sec:bif} previously.
Instead, we observe a reorganization of the convection cells through local merging events.
Figure~\ref{fig:tmid} shows the temperature profile located at the mid-height of the fluid domain, $\theta(x,z=\overline{h}(t)/2,t)$.
At early times, as shown in Figure~\ref{fig:tmid}(b), the temperature profile is initially purely diffusive and uniform, $\theta(z=\overline{h}/2)=(1-\theta_M)/2\simeq1/2$.
The first network of steady convective cells eventually appears and we again observe that the typical horizontal wavelength remains constant after the first nonlinear saturation of the convection instability.
The secondary instability, which involved a horizontal mean flow in section~\ref{sec:bif}, now appears to be more local since the system is not laterally confined.
Interestingly, these local transitions tend to propagate horizontally to neighbouring cells in a percolation process.
This is indicated in Figure~\ref{fig:tmid}(b) by the inclined dashed lines.
The typical speed of propagation of this defect in the convection cells lattice, estimated directly from the slope of the dashed lines in Figure~\ref{fig:tmid}(b), is approximately the same as the fluid vertical velocity.
Each cell is therefore destabilized after approximately one turnover time.

In addition, the thermal plumes are clearly oscillatory just after the bifurcation but eventually stabilize and become steady after a short transient.
This observation is especially interesting since the Rayleigh number of the system is continuously increasing with time.
One might therefore expect the dynamics to become more and more complex, transiting from periodic to chaotic and eventually turbulent solutions as it is the case in classical RB convection without topography. 
This transition from oscillatory convection to steady convection clearly shows the stabilizing effect that the topography exerts on the flow, locking two counter-rotating convection rolls inside each cavity.
At later times and higher Rayleigh numbers, shown in Figure~\ref{fig:tmid}(a), although the stabilization of the thermal plumes by the topography is still observed, significant temporal fluctuations are nevertheless visible, indicating that the convective cells will eventually transit to more chaotic behaviours.
The inevitable transitions between steady, periodic and chaotic solutions observed in classical RB \citep{gollub_benson_1980,curry_herring_loncaric_orszag_1984,goldhirsch_pelz_orszag_1989} are therefore probably just delayed by the presence of the topography, but will eventually reappear at much larger Rayleigh numbers.
This conclusion remains speculative at this stage since this particular simulation is limited to $Ra_e<10^8$.
It is nevertheless reasonable to expect a different, potentially reduced, interaction between the topography and the underlying flow in the fully developed turbulent regime.

A final interesting observation concerns the clear asymmetry between rising and sinking plumes.
Sinking plumes are extremely stable and do not seem to move horizontally, apart from the sudden transitions associated with the reorganisation of the convective cells.
As seen in Figure~\ref{fig:visus}, cold plumes are generated by the merging of two boundary layers descending along the topography, leading to the formation of a high curvature cusp.
This cusp is therefore protected by the continuous supply of cold fluid generated by the melting of the neighbouring dome by hot rising fluid.
The sinking plumes are therefore always found to be emitted by the cusps.
In contrast, rising plumes tend to slowly drift horizontally until they eventually collide with an adjacent sinking plume, leading to destabilisation of both convection rolls.
The reason of this drift is probably associated with the baroclinic gravity currents, infinitesimal at low Rayleigh numbers and small topography as in section~\ref{sec:crit}, but much stronger at large effective Rayleigh numbers and for finite topography slopes.
Once a rising thermal plume slightly moves horizontally, it is continuously dragged by the topographic current until a merger occurs.
The competition between thermal convection driven by unstable bulk temperature gradients and gravity currents driven by a baroclinic forcing close to an inclined slope is interesting in itself, although we postpone the study of its detailed dynamics to future studies.

\section{Statistical description\label{sec:st}}

We now describe the evolution of the convection and of the topography in a more quantitative way by systematically varying the Stefan number and measuring the averaged response in the interface $h(x,t)$ and in the heat flux $Q_W$.
In order to reach the large Stefan number regime, for which the solid melts at a much slower rate, we reduce the Rayleigh number to $Ra=10^7$, the other parameters being showed in Table~\ref{tab:one}, case D.

\subsection{Melting velocity}

We now consider the effect of varying the Stefan number on the dynamics of the convective flow and interface.
The parameters are the same as previously but we now vary the Stefan number from $St=2\times10^{-2}$ to $St=50$.
The main consequence of increasing the Stefan number is to increase the timescale separation between the turnover time of the convective cells and the typical timescale of evolution of the topography.
As $St$ increases, it takes much more time for a given set of convection rolls to form or alter a topography due to the larger amount of latent heat necessary to do so.

\begin{figure}
   \vspace{6mm}
   \centering
   \includegraphics[width=0.53\textwidth]{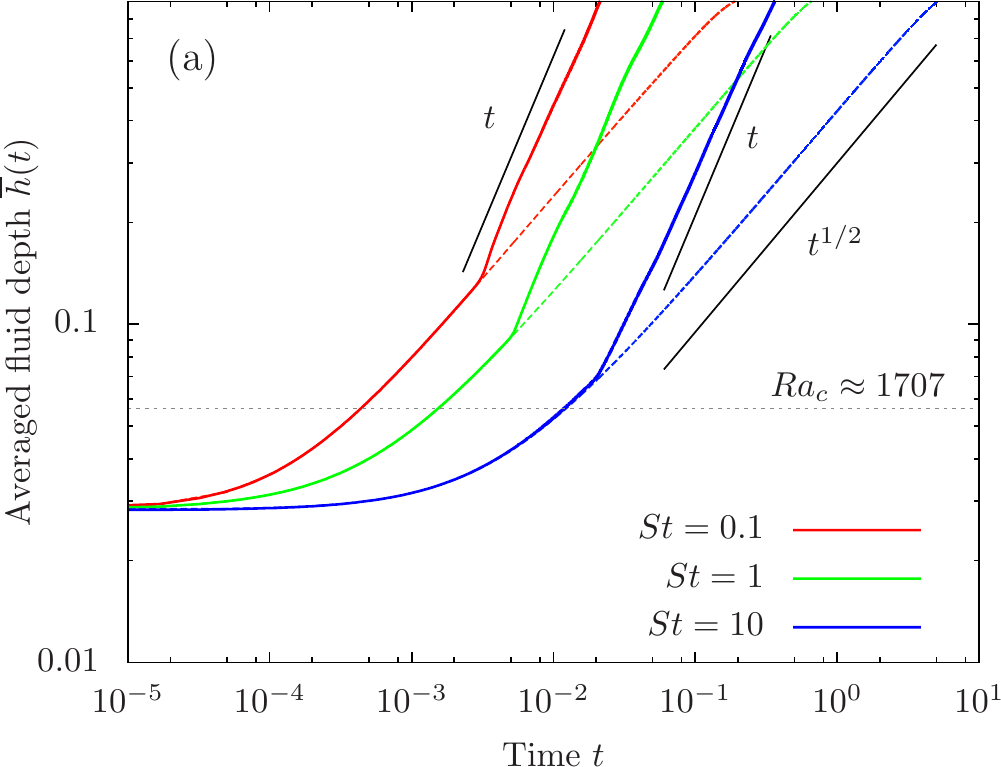}
   \hspace{2mm}
   \includegraphics[width=0.44\textwidth]{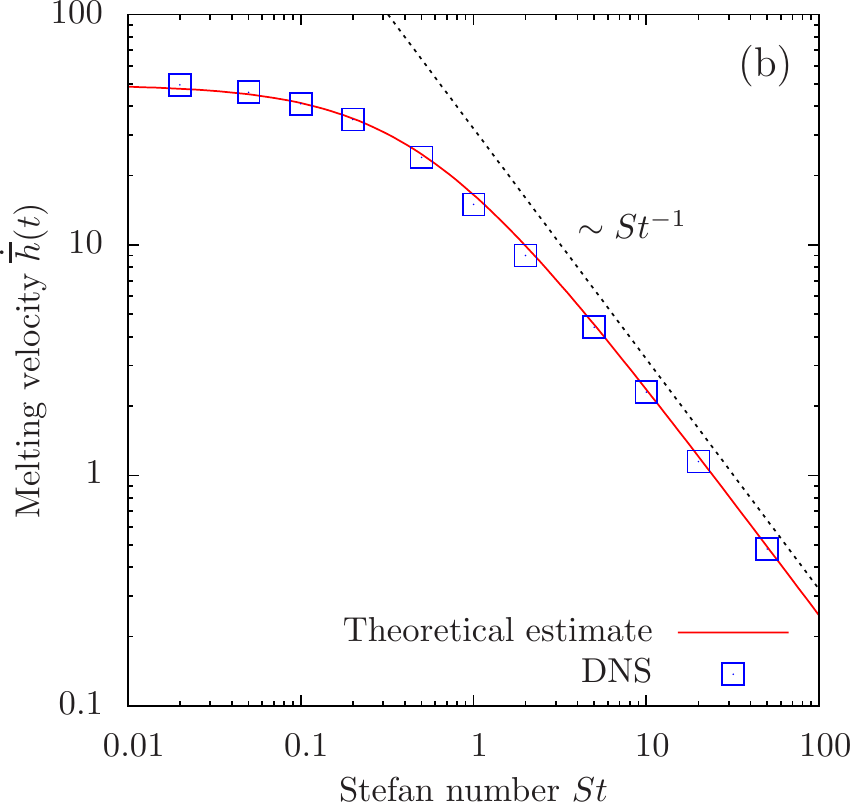}
   \caption{(a) Time evolution of the horizontally averaged fluid depth $\overline{h}$ for different Stefan numbers. The two scalings $\overline{h}\sim t$ and $\overline{h}\sim t^{1/2}$ are shown as continuous black lines. The two horizontal dotted lines correspond to the critical heights above which convection sets in, estimated from $Ra_c\approx1707$. (b) Melting velocity $\dot{\overline{h}}$ for different Stefan numbers. The theoretical estimate is derived from equation~\eqref{eq:hdot} using $\gamma\approx0.115$ and $\beta=1/3$.\label{fig:hbar}
   }
\end{figure}

This can be seen in Figure~\ref{fig:hbar}(a) where the averaged fluid depth is shown versus time for three different Stefan numbers.
The dotted lines show the purely diffusive solution in the absence of motions in the fluid phase (\textit{i.e.} for $Ra_e<Ra_c$ at all times).
These purely diffusive solutions all display the scaling $\overline{h}\sim t^{1/2}$ as expected for diffusive Stefan problems \citep{Vasil2011}.
One observe a departure from this prediction which marks the onset of convection.
The larger the Stefan number, the longer it takes to reach the threshold of convection.
However, all cases follow the scaling $\overline{h}\sim t$ after the onset of convection.
The prefactor however depends on the Stefan number, as shown in Figure~\ref{fig:hbar}(b).
The melting velocity, obtained by a best fit of the previous linear law, is plotted against the Stefan number.
The melting velocity seems to be independent of the Stefan number for low values and scales as $St^{-1}$ for large values.

These behaviours can be understood by simple energetic arguments.
By integrating equation~\eqref{eq:tempeq_adim} over the whole volume $\mathcal{V}$, we find the following relation
\begin{equation}
\label{eq:cons}
\frac{d}{dt}\int_{\mathcal{V}} \Big[\theta+St \ p(\phi)\Big] \textrm{d}\mathcal{V} = \int_{\textrm{upper}}\frac{\partial \theta}{\partial z} \textrm{d}S - \int_{\textrm{lower}}\frac{\partial \theta}{\partial z} \textrm{d}S \ ,
\end{equation}
where the left-hand side corresponds to the total rate of change of the enthalpy in the system whereas the right-hand side corresponds to the heat fluxes entering and leaving the domain.
This conservation constraint must be exactly satisfied at all times during the simulations.
Since we work with the temperature as a variable with the latent heat being viewed as an external forcing term, the enthalpy is not explicitly conserved by our scheme.
We therefore have to check \textit{a posteriori} that the enthalpy is indeed conserved in our system.
We typically observe a relative error in the total enthalpy of the system of the order of $1\%$ at the final time of the simulations when all the solid has melted.

Equation~\eqref{eq:cons} can also be used to estimate the rate of change of the average fluid height $\overline{h}(t)$.
The internal heat associated with the solid can be neglected in first approximation since we consider the limit where $\theta_M\ll1$.
In addition, assuming that the fluid layer is fully convective and behaves as in classical RB convection, its average temperature can be approximated by $1/2$.
This leads to the following relation (ignoring heat losses from the quasi-isothermal solid above)
\begin{equation}
\label{eq:tb}
\left(\frac12+St\right)\frac{dV_f}{dt} = -\int_{\textrm{lower}}\frac{\partial \theta}{\partial z} \textrm{d}S\approx\lambda\frac{Nu}{\overline{h}} \ ,
\end{equation}
where the heat flux from the lower boundary has been replaced by the Nusselt number $Nu$ as defined in equation~\eqref{eq:nuss} (and we have assumed that $\theta_M\ll1$).
In equation~\eqref{eq:tb}, the volume of fluid
\begin{equation}
    V_f=\int_{\mathcal{V}} p(\phi) \textrm{d}\mathcal{V}
\end{equation}
can be approximated by $\overline{h}\lambda$ and the Nusselt number can be replaced by the usual scaling law involving the effective Rayleigh number of the form $Nu\sim \gamma Ra_e^{\beta}$ (see Section~\ref{sec:hflux} for a more detailed discussion of the heat flux) leading to
\begin{equation}
\left(\frac12+St\right) \frac{d\overline{h}}{dt} \approx \frac{\gamma Ra_e^{\beta}}{\overline{h}}\approx\gamma Ra^{\beta}\overline{h}^{3\beta-1} \ ,
\end{equation}
where we have again assumed that $\theta_M\ll1$.
The solution to this equation reads
\begin{equation}
\label{eq:hdot}
    \overline{h}(t)\approx \left[h_0^{2-3\beta}+\frac{(2-3\beta)\gamma Ra^{\beta}}{1/2+St}t\right]^{1/(2-3\beta)} \ ,
\end{equation}
where $h_0=h(t=0)$.
Using the typical value of $\beta=1/3$ \citep{grossmann_lohse_2000}, we find that $\overline{h}\sim t$ which is indeed recovered by our simulations.
Note that assuming that $\beta=1/4$ leads to  $\overline{h}\sim t^{4/5}$ which is also in reasonable agreement with our simulations.
In addition, assuming that $\beta=1/3$, the melting velocity $\dot{\overline{h}}$ can be estimated for different Stefan numbers from equation~\eqref{eq:hdot} and compared with the numerical results.
The only adjusting parameter is the prefactor $\gamma$ linking the Nusselt number with the Rayleigh number.
We show in Figure~\ref{fig:hbar} the best fit with our numerical data which gives $\gamma\approx0.115$.
The agreement is very good over nearly four decades of Stefan numbers.
In particular, we recover the fact the melting velocity behaves like $St^{-1}$ in the limit of large Stefan numbers.

\subsection{Horizontal and vertical scales of the topography}

\begin{figure}
   \vspace{5mm}
   \centering
   \includegraphics[width=0.455\textwidth]{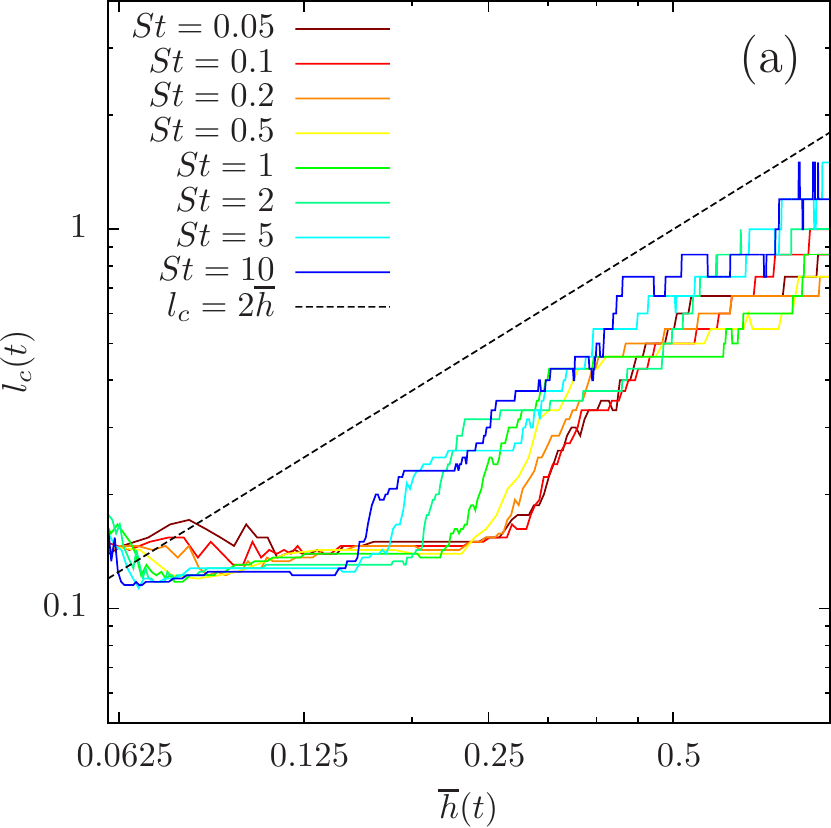}
   \hfill
   \includegraphics[width=0.48\textwidth]{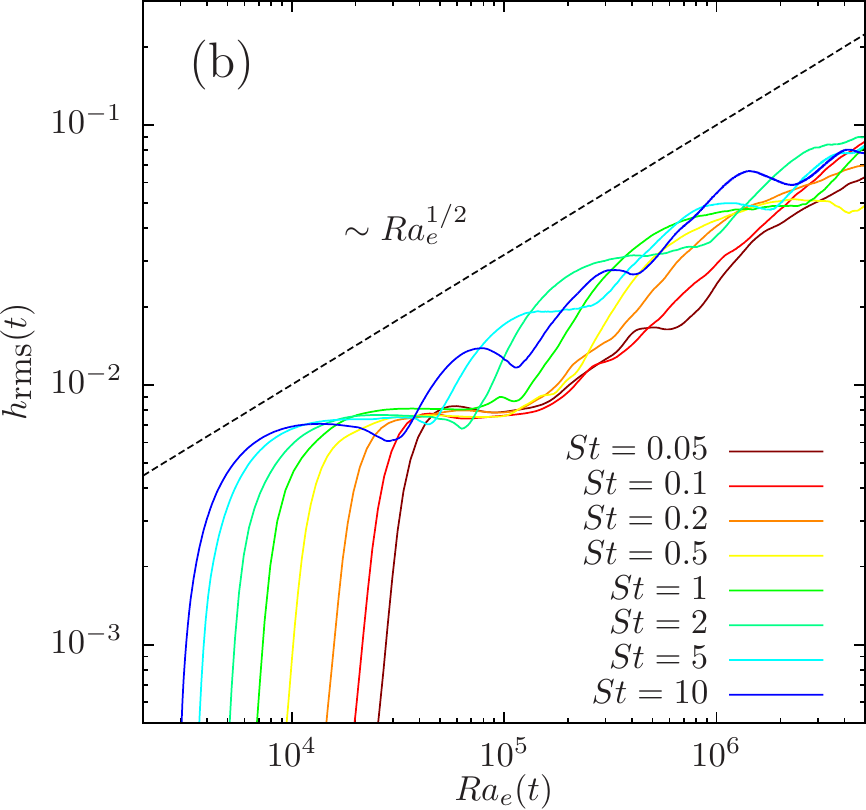}
   \caption{(a) Typical horizontal extent of the cavities as a function of the averaged fluid depth. The dashed line corresponds to the limit case where each convective roll has a unit aspect ratio. (b) Root-mean-square value of the topography defined by equation~\eqref{eq:rmsh} as a function of the effective Rayleigh number and for different Stefan numbers. The scaling $Ra_e^{1/2}$ is shown for reference. \label{fig:lc}}
\end{figure}

Let us now discuss some properties of the topography as times evolves.
We can track the number of local minima $N_{\textrm{min}}(t)$ of the function $h(x,t)$ as a function of time.
The typical length of the cavities generated by flow can be estimated as $l_c(t)=\lambda/N_{\textrm{min}}(t)$ where $\lambda$ is the aspect ratio of the numerical domain.
This length-scale is plotted in Figure~\ref{fig:lc} for different Stefan numbers as a function of the average fluid depth $\overline{h}(t)$.
As expected, the typical size of the cavities grows with time.
Additionally, as seen in Figure~\ref{fig:visus}, each cavity contains two convective rolls, each having an opposite circulation.
For classical RB convection, the convective rolls typically have a unit aspect ratio (which is related to the fact the critical wave number is approximately $k\approx\pi$).
This corresponds to the typical relation $l_c(t)\approx2\overline{h}(t)$, which we also plot in Figure~\ref{fig:lc}.
All the curves remain below this curve, showing that our convective rolls are always slightly elongated in the vertical direction despite the successive dynamical transitions that eventually destabilize them.
Note that convective cells tend to be less elongated (\textit{i.e.} our results get closer to the prediction $l_c=2\overline{h}$) as the Stefan number increases.
This is a direct consequence of the timescale separation typical of the large Stefan number regime, for which the flow can quickly bifurcate to a new, more unstable, set of convective rolls without any significant change in the average fluid depth. 

Finally, it is interesting to consider the typical amplitude of the topography.
One can compute the root-mean-square depth as 
\begin{equation}
\label{eq:rmsh}
 h_{\textrm{rms}}(t)=\sqrt{\overline{\left(h(x,t)-\overline{h}(t)\right)^2}} \ .   
\end{equation}
Figure~\ref{fig:lc} shows this quantity as a function of the effective Rayleigh number of the fluid layer.
Interestingly, there is little dependence with the Stefan number, except close to onset where the saturation of convective occurs later for small Stefan numbers.
The typical amplitude of the topography, as measured by $h_\textrm{rms}$, seems to depend mostly on the effective Rayleigh number of the layer, following an approximate scaling of $Ra_e^{1/2}$.
Note that the fluctuations of our results, a consequence of the dynamical transitions discussed above, and the relatively small variation in effective Rayleigh numbers, are limiting us from getting a conclusive scaling.
It is however interesting to note that the local Reynolds number of RB convection typically scales as $Ra^{1/2}$ \citep{grossmann_lohse_2000}, so that there might be a link between the amplitude of the topography and the Reynolds number of the underlying flow, and thus the thickness of the boundary layers developing along the topography.

\subsection{Effect of the topography on the heat flux\label{sec:hflux}}

The previous section has showed the non-trivial back-reaction that the topography imprints on the convective flow.
The effect of non-uniform boundary conditions on the heat flux in a Rayleigh-B\'enard system has a long history.
Roughness of the horizontal plates is an obvious candidate to trigger boundary layer instabilities possibly leading to an enhancement of the heat flux \citep{Ciliberto1999} and possibly to the so-called ultimate regime predicted by Kraichnan \citep{Kraichnan1962,Roche2010}.
The wave-length and typical amplitude of our topography is however much larger than the typical roughness used in experiments \citep{rusa_2018}.
Note also that roughness does not always lead to a heat transfer enhancement \citep{zhang_sun_bao_zhou_2018}.
In the present case, the horizontal wavelength of the topography is precisely that of the most unstable wavelength of the idealized Rayleigh-B\'enard problem in the absence of topography, a situation sometimes referred as to the resonant case \citep{Kelly1978,Bhatt1991,Weiss2014}.

\begin{figure}
   \vspace{5mm}
   \centering
   \includegraphics[width=0.7\textwidth]{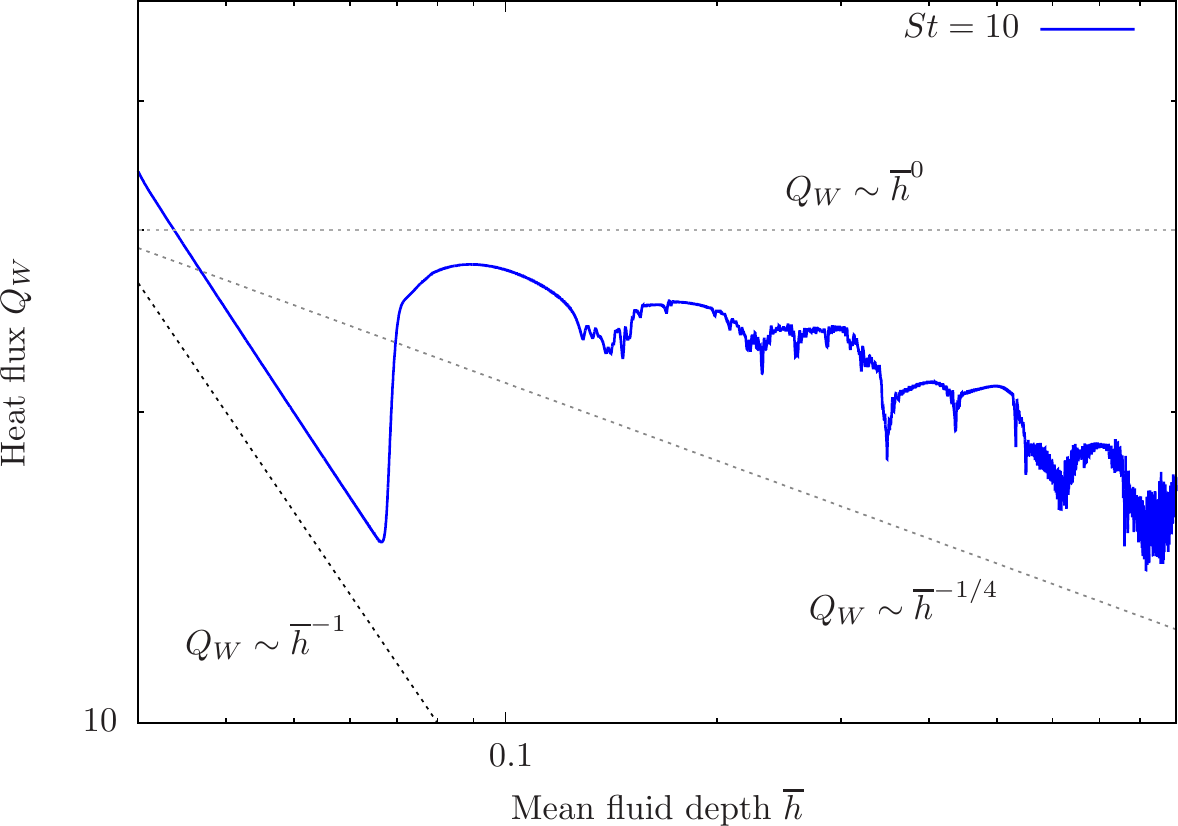}
   \caption{Heat flux averaged over the bottom surface $Q_W$, as defined in equation~\eqref{eq:thf}, as a function of the averaged fluid depth $\overline{h}(t)$. The results are only showed for the case $St=10$ for clarity. The dotted lines correspond to various power law scalings.\label{fig:qw}}
\end{figure}

In this section, we consider the heat flux at the bottom boundary $Q_W$ defined by equation~\eqref{eq:thf}.
We show the evolution with time of this heat flux for $St=10$ and $Ra=10^7$ as a function the average fluid depth $\overline{h}$ in Figure~\ref{fig:qw}.
Before the onset of convection, the purely diffusive heat flux is simply given by $Q_W\approx(1-\theta_M)/\overline{h}$ that is indeed observed initially.
After convection sets in, we observe a rapid increase of heat flux associated with the nonlinear overshoot of the instability.
The heat flux then tends to decrease with time, but we also observe a succession of plateaus characterized by an approximately constant heat flux $Q_W$, separated by sudden decays.
The plateaus correspond to the quasi-steady phases where the convection is locked inside the topography, whereas the sudden decays correspond to the secondary bifurcation where the mean flow is disrupting the convection rolls and inhibits the heat flux across the fluid layer.
Starting from the classical relation $Nu\sim \gamma Ra_e^{\beta}$, where $Nu$ is the Nusselt number defined by equation~\eqref{eq:nuss}, leads to the relation $Q_W\sim\overline{h}^{3\beta-1}$ so that for $\beta<1/3$, the convective heat flux is indeed a decreasing function of the fluid height, all other parameters being fixed, whereas it is independent of the fluid height when $\beta=1/3$.
These different scalings are shown in Figure~\ref{fig:qw} for reference.

We now consider the problematic question of the normalization of this heat flux.
In classical RB convection, the diffusive flux across a plane-layer domain is trivially derived from the solution of the purely diffusive heat equation.
In our case however, the diffusive flux is not trivial since the topology of the fluid domain is fully two-dimensional and of finite amplitude.
Formally, one should therefore solve the heat equation in order to know the diffusive heat flux across the layer.
This refinement has negligible consequence when the topography is of very small amplitude but this is not the case here, where the topography is of comparable order with the fluid depth.
We therefore derive below a second-order correction of the diffusive heat flux at the bottom boundary.

We consider the purely diffusive case of a fluid layer heated from below ($\theta=1$ in $z=0$), with the upper surface at temperature $\theta=\theta_M$ located at
\begin{equation}
z = h(x,t) = \overline{h}(t) \left( 1+ \varepsilon \cos{k x} \right) \ , \label{h_exp}
\end{equation}
where $\varepsilon \ll 1$, $\overline{h}$ is the mean height given by equation \eqref{meanheight}, and $k$ is the wave number of the topography. 
We assume that the evolution of $\overline{h}(t)$ is much slower than the diffusion (which is formally justified in the large Stefan limit, see~\cite{Vasil2011}).
Therefore, we note $\overline{h} = h_0$, $\Delta \theta=1-\theta_M$, and we look only for stationary solutions of the diffusion equation
\begin{equation}
    \nabla^2 \theta = 0, \qquad z \in [0,h(x)] \ , \label{laplace}
\end{equation}
with boundary conditions
\begin{eqnarray}
\theta(x,0) & = & 1\\
\theta(x,h(x)) & = & \theta_M \ . \label{upper_bound}
\end{eqnarray}
We expand $\theta$ in power series of $\varepsilon$
\begin{equation}
\theta(x,z) = \theta_0(x,z) + \varepsilon \theta_1(x,z) + \varepsilon^2 \theta_2(x,z) + \cdots
\end{equation}
and solve for equation \eqref{laplace} at each order in $\varepsilon$. 
After some algebra (Cf. Appendix \ref{sec:appC}), we obtain at second-order
\begin{equation}
Q_D  =-\frac{1}{\lambda}\int_0^{\lambda}\frac{\partial\theta}{\partial z}(x,0)\text{d}x= 
\frac{\Delta \theta}{h_0}
+\varepsilon^2 \frac{k \Delta \theta}{2} \coth{k h_0} \ .
\label{eq:th_hf2}
\end{equation}
On the other hand, the area of the topography per unit horizontal length $A$ (a length per unit length in our 2D geometry) can be computed using equation \eqref{h_exp}
\begin{equation}
A = \frac{1}{\lambda} \int_0^\lambda {\sqrt{1+\left(\frac{\partial h(x,t)}{\partial x}\right)^2}}dx
= \frac{1}{\lambda} \int_0^\lambda {\sqrt{1+\left(-kh_0 \varepsilon \sin{kx}\right)^2}}dx \ .
\end{equation}
At order $\varepsilon^2$, we obtain
\begin{equation}
\Delta A \equiv A - 1 = \frac{1}{4} k^2 h_0^2 \varepsilon^2 \ .
\end{equation}
Finally, the diffusion heat flux can be expressed according to this surface area increase
\begin{equation}
Q_D = \frac{\Delta \theta}{h_0} 
\left(
1 + \frac{2 \Delta A}{k h_0} \coth{k h_0} \ .
\right)
\end{equation}
The typical wavelength in our simulations being of the order of $2h_0$, we can estimate this diffusion flux by writing $k \simeq \pi/h_0$ leading to
\begin{equation}
Q_D \simeq \frac{\Delta \theta}{h_0} 
\left(
1 + \frac{2\Delta A}{\pi}
\right) \ .
\end{equation}

We can now rescale the heat flux through the fluid layer $Q_W$ at each time knowing the averaged fluid depth $\overline{h}$ and the area increase $\Delta A$ computed numerically at each time step.
The diffusive heat flux through the fluid layer is approximately
\begin{equation}
\label{eq:diff*}
    Q_D(t)=\frac{1-\theta_M}{\overline{h}(t)}\left(1+\frac{2\Delta A(t)}{\pi}\right)
\end{equation}
and the Nusselt number is finally defined as the ratio between to total heat flux~\eqref{eq:thf} and the diffusive flux estimated using equation~\eqref{eq:diff*}.

\begin{figure}
   \vspace{5mm}
   \centering
   \includegraphics[width=0.8\textwidth]{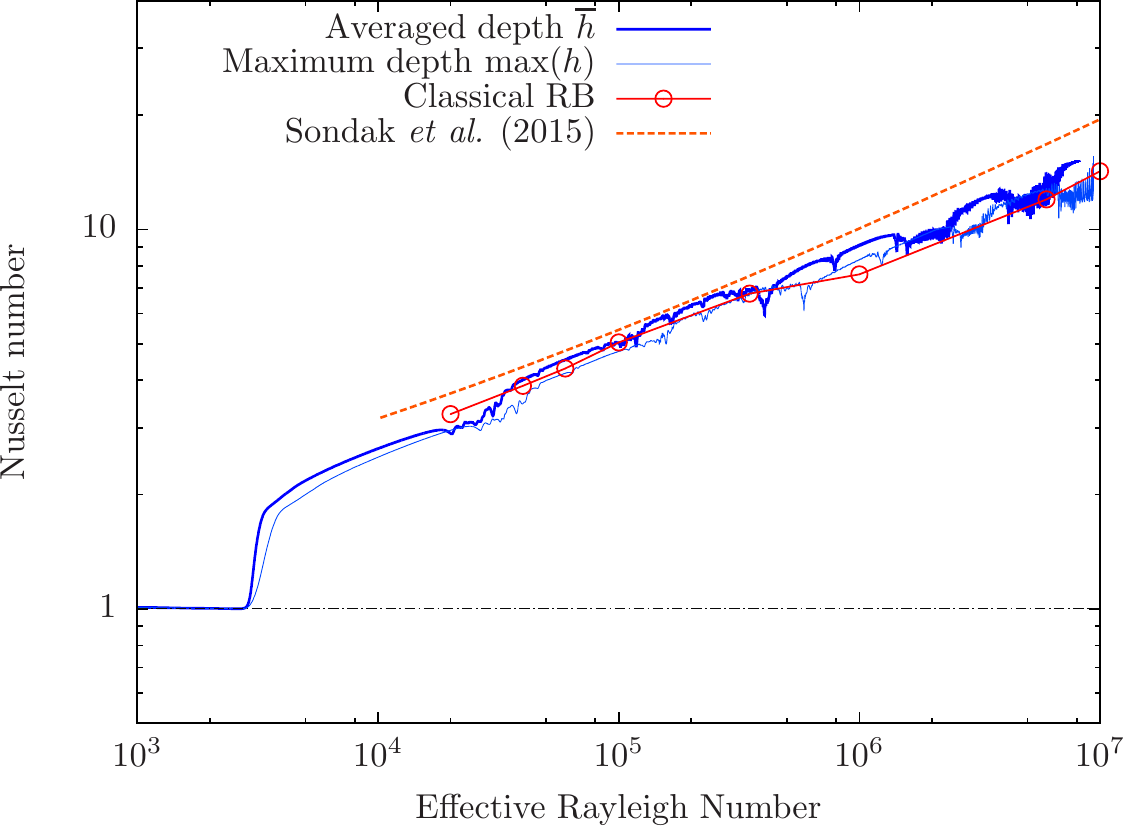}
   \caption{Nusselt number as a function of the effective Rayleigh number $Ra_e(t)$ for a Stefan number of $St=10$. The empty symbols correspond to the classical Rayleigh-B\'enard case whereas the dashed line corresponds to the optimum steady solution of \cite{sondak_smith_waleffe_2015}. The thin line is shown for indication and is obtained by computing the effective Rayleigh number on the maximum fluid height instead on its horizontally-averaged value.\label{fig:nussf}}
\end{figure}

The result of this normalization is shown in Figure~\ref{fig:nussf}.
For the case $St=10$, we show the time evolution of the effective Nusselt number as a function of the effective Rayleigh number.
For reference, we also show some typical values obtained for classical Rayleigh number (each point corresponding in that case to the time average of a single simulation at fixed Rayleigh number).
We also indicate the results of \cite{sondak_smith_waleffe_2015} which correspond to the optimal heat transfer for a 2D steady solution giving $Nu\approx0.125 Ra^{0.31}$ \footnote{In \cite{sondak_smith_waleffe_2015}, the best fit is actually $Nu\approx0.115Ra^{0.31}$ but corresponds to a Prandtl number of 7. They obtained a slightly larger prefactor for $\sigma=1$ but the same power law.}.
Interestingly, although our simulation departs significantly from classical Rayleigh-B\'enard, our renormalization shows that the Nusselt number follows that of classical RB convection in a quasi-static manner.
This is of course only true at large Stefan numbers.
For lower Stefan numbers (not shown), the curves are much more erratic and no clear trend can be derived.
There are however significant differences between our case and the predictions specific to RB.
Although the exponent is not significantly altered by the presence of the melting interface, the prefactor appears larger than what it is for regular RB.
This is marginally true at low Rayleigh numbers ($Ra_e<10^6$) but quite clear at higher Rayleigh numbers.
This can be attributed to the back-reaction of the topography on the convective rolls, which appears to be a stabilizing effect by delaying the transition to periodic convection.
In 2D, this transition typically occurs around $Ra\approx10^5$ and reduces the Nusselt number, as can be seen in Figure~\ref{fig:nussf}.
The presence of the topography induced by the convective flow itself seems to favor stable quasi-steady rolls as opposed to oscillatory ones.
This leads to an increase in heat flux when compared to classical RB and is closer to the optimal solution of \cite{sondak_smith_waleffe_2015}, derived assuming steady laminar solutions.
This marginal increase in the Nusselt number was also recently reported in the independent study by \cite{Babak2018}, both in two and three dimensions, although for a Prandtl number of $10$.
Note finally that although we carefully normalized the heat flux, our choice of Rayleigh number is rather arbitrary.
One could argue for example that the effective Rayleigh number based on the averaged fluid depth is barely relevant and that only the maximum depth where the Rayleigh number is effectively maximum matter for the heat flux.
The results corresponding to this particular choice is shown in Figure~\ref{fig:nussf} as the thin line.
Although it does reduce the overall heat flux for a given Rayleigh number, our conclusions drawn above remain qualitatively valid.
Since we are limited in the maximum value for our effective Rayleigh number (a consequence of the finite vertical extent of our numerical domain), it is not clear how the topography affects the heat flux in the fully chaotic regime reached at much higher Rayleigh numbers.

\section{Conclusion\label{sec:conclu}}

Numerical simulations of Rayleigh-B\'enard convection in two dimensions with an upper melting boundary have been performed.
We have shown that the fact that the upper boundary dynamically becomes non-planar has interesting consequences on the development of convection in the fluid layer.
The onset of convection becomes imperfect due to baroclinic effects close to the topography so that it is difficult to study the transition between a purely diffusive regime and thermal convection, even when the Stefan number is large.
The initial saturation of the instability leads to steady convective rolls carving a topography with the same wavelength.
As the fluid depth increases and when the flow is laterally confined, the steady rolls eventually feed a mean horizontal shear flow, as observed in supercritical RB convection, which disrupts convection until a new array of convective rolls grows with an aspect ratio close to unity.
For large horizontal aspect ratios, the transition is replaced by local merging events propagating to neighbouring cells in a percolation process.
Finally, at higher Rayleigh numbers, we observe that the convection rolls remain locked into the topography, delaying bifurcations to periodic and chaotic orbits, and effectively increasing the heat flux compared to classical RB convection.

Many aspects of this apparently simple system remain to be explored.
We focused in this paper on the particular case $\theta_M\rightarrow0$.
It is however obvious that the system can reach a quasi-steady equilibrium with both liquid and solid phases present when $0<\theta_M<1$.
In that case, the solid is cooled from above, effectively balancing the heat flux brought by thermal convection in the liquid phase below.
Depending on the values for $\theta_M$, $Ra$ and $St$, this regime is expected to lead to interesting dynamics which we are currently exploring.
The Prandtl number was also fixed to be unity for simplicity but it is well known that classical RB convection crucially depends on this parameter and we expect our system to be the same.
It is also worth recalling that liquid metals typically have very low Prandtl numbers, for which we expect a different melting or solidifying dynamics.
Finally, while it would be very difficult to generalize our approach to variable densities between the solid and liquid phases, a natural extension involving non-uniform thermal diffusivities is nevertheless possible \citep{Almgren1999}.

Based on the phase-field method, our approach is relatively simple to implement in existing numerical codes capable of solving the usual Boussinesq equations.
As of now, it remains numerically expensive due to the fact that some diffusive terms are solved explicitly.
This could easily be improved by considering a linearized version of the last term on the right-hand size of equation~\eqref{eq:temp}, allowing for a fully-implicit coupling between the temperature and the phase field equations.
This would allow us to consider similar problems in three-dimensions, thus extending the early experimental works by \cite{davis_muller_dietsche_1984} and the recent numerical study of \cite{Babak2018}.
Some of the results discussed in this paper might not be relevant to the three-dimensional case since the stability of 3D convection patterns are notoriously different from their 2D equivalent.
The effects of an upper melting boundary on the development of 3D convection cells remain to be fully explored.

The framework developed in this paper could finally be used to study other free-boundary problems.
The dynamical creation of non-trivial topographies by dissolution \citep{claudin_duran_andreotti_2017} or erosion \citep{Matthew2013}
are also accessible using the current approach.
The Stefan boundary condition which depends on the temperature gradients can be generalized to incorporate gradients of concentration or tangential velocity.
Several academic configurations could be therefore revisited using a continuous interface approach such as the phase field model coupled with the Navier--Stokes equations.

\vspace{1cm}
\textbf{Acknowledgments.} We gratefully acknowledge the computational hours provided on Turing and Ada (Projects No. A0020407543 and A0040407543) of IDRIS.
This work was granted access (Project No. 017B020) to the HPC resources of Aix-Marseille Universit\'e financed by the project Equip@Meso (ANR-10-EQPX-29-01) of the program ``Investissements d'Avenir'' supervised by the Agence Nationale de la Recherche.
We gratefully acknowledge financial support from the Programme
National de Plan\'etologie (PNP) of the Institut National des Sciences de l’Univers (INSU, CNRS).
We have finally benefited from many discussions with Geoffrey Vasil.

\appendix
\section{Numerical convergence with the phase-field parameters\label{sec:appA}}

The objective of this section is to show how the parameters introduced by the phase-field formulation are chosen in order to recover the original Stefan problem as defined by equations~\eqref{eq:stefan1}-\eqref{eq:st2}.

\subsection{One-dimensional case without fluid motions\label{sec:A1}}

We consider a simple one-dimensional Stefan problem.
A liquid phase of a pure material initially occupies a region $[0,x_i]$ while the solid phase occupies the region $[x_i,1]$.
The dimensionless temperature is imposed to be $\theta=1$ at $x=0$ and $\theta=0$ at $x=1$.
The interface is initially located at $x_i(t=0)=1/5$ and the melting temperature is $\theta_M=1/5$.
The initial temperature profile is given by $\theta(x,t\!=\!0)=(e^{-\beta(x-1)}\!-\!1)/(e^{\beta}\!-\!1)$
where $\beta\approx8.041$.
The Stefan problem \eqref{eq:stefan1}-\eqref{eq:st2} is then solved for a fixed Stefan number of $St=1$.
The expected steady state, which is recovered by all simulations discussed in this section, corresponds to a linear temperature profile $\theta(x,t\rightarrow\infty)=1-x$ and $x_i(t\rightarrow\infty)=4/5$.
In the following, we focus on the more interesting transient phase and compare different numerical methods in their ability to predict the position of the front and the temperature profile at $t=1/4$.

\begin{figure}
   \vspace{5mm}
   \centering
   \includegraphics[width=0.47\textwidth]{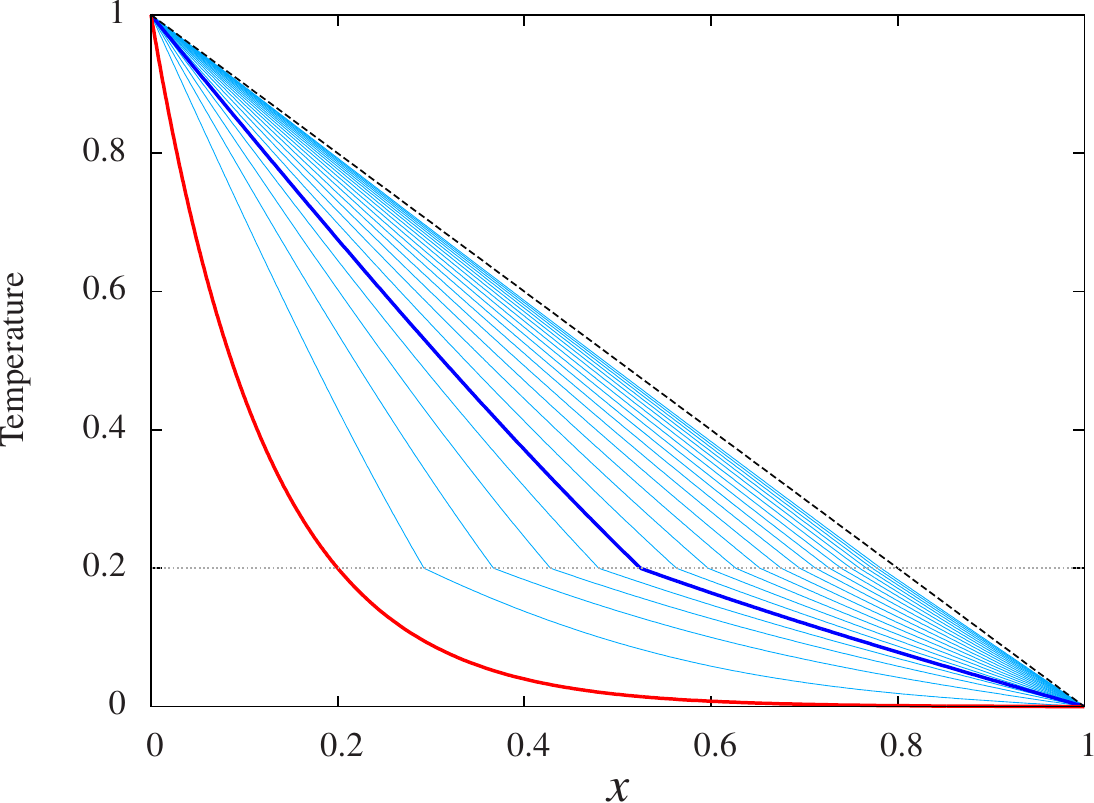}
   \hspace{2mm}
   \includegraphics[width=0.47\textwidth]{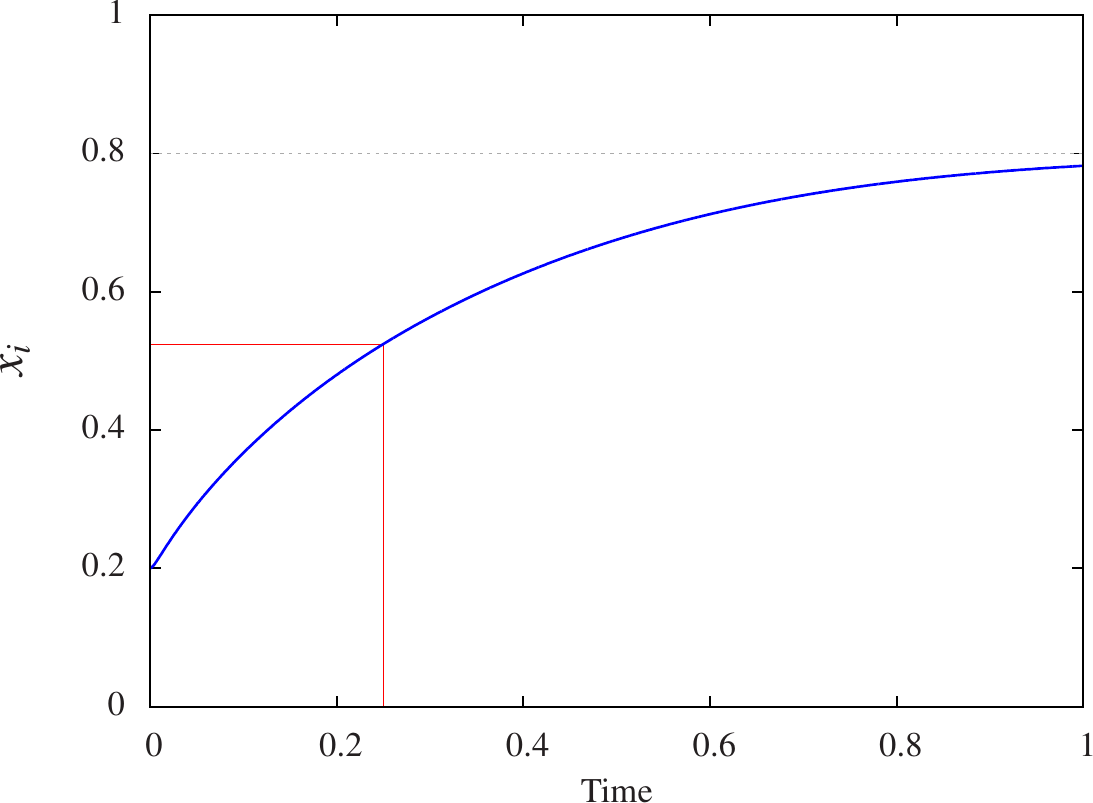}
   \caption{Reference solution obtained by the mapping method.
   Left: time evolution of the temperature profile. The red curve corresponds to the initial condition whereas the blue curve corresponds to the reference time $t=1/4$ used for comparison. Right: time evolution of the interface position.\label{fig:Aref}}
\end{figure}

The reference solution is obtained by a mapping method where the unsteady liquid domain $x\in[0,x_i(t)]$ is transformed into a steady domain $\zeta\in[0,1]$ whereas the solid domain $x\in[x_i(t),1]$ is transformed into $\zeta\in[1,2]$.
This is achieved by using the following change of spatial variable
\begin{equation}
\zeta(t)=\left\{
\begin{array}{c}
\displaystyle{x/x_i(t)} \quad \textrm{in the liquid domain} \\
\displaystyle{(x-2x_i(t)+1)/(1-x_i(t))} \quad \textrm{in the solid domain}
\end{array}
\right. \
\end{equation}
where $\zeta=1$ corresponds to the interface at all times.
In each domain, the spatial derivative in $\zeta$ are discretized using fourth-order finite differences whereas the time-stepping is performed with a third-order explicit Adams-Bashforth scheme.
We use the same number of grid-points $N=256$ in each domain.
The temporal evolution of the temperature profile and of the position of the interface are shown in Figure~\ref{fig:Aref}.

\begin{figure}
   \vspace{5mm}
   \centering
   \includegraphics[width=0.47\textwidth]{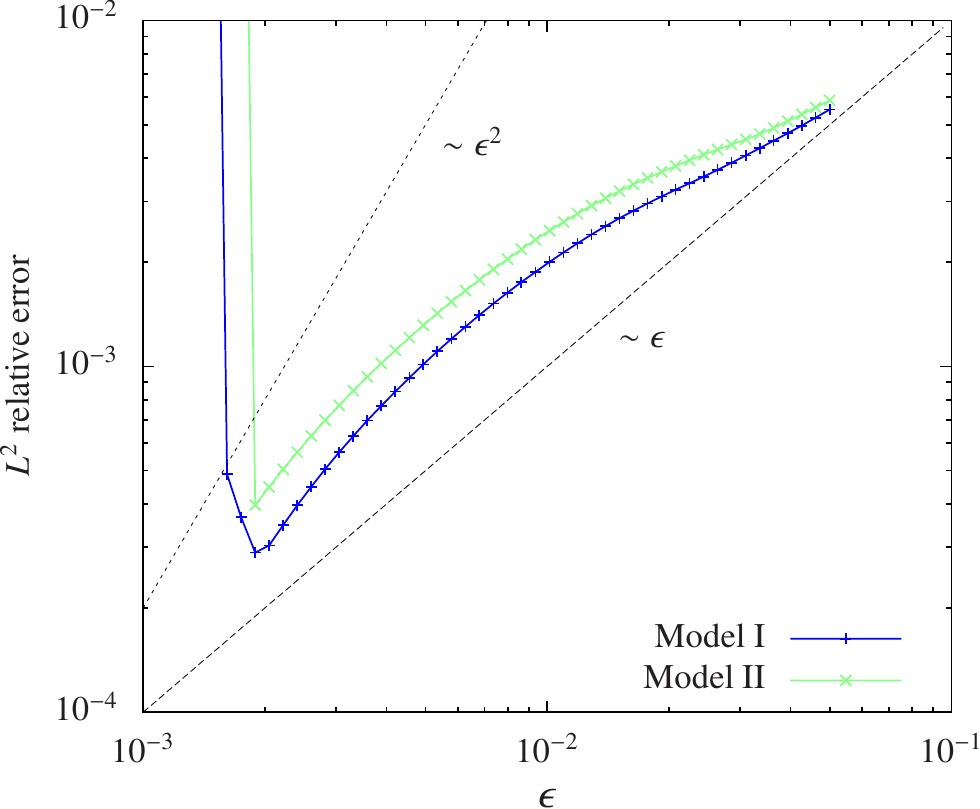}
   \hspace{2mm}
   \includegraphics[width=0.47\textwidth]{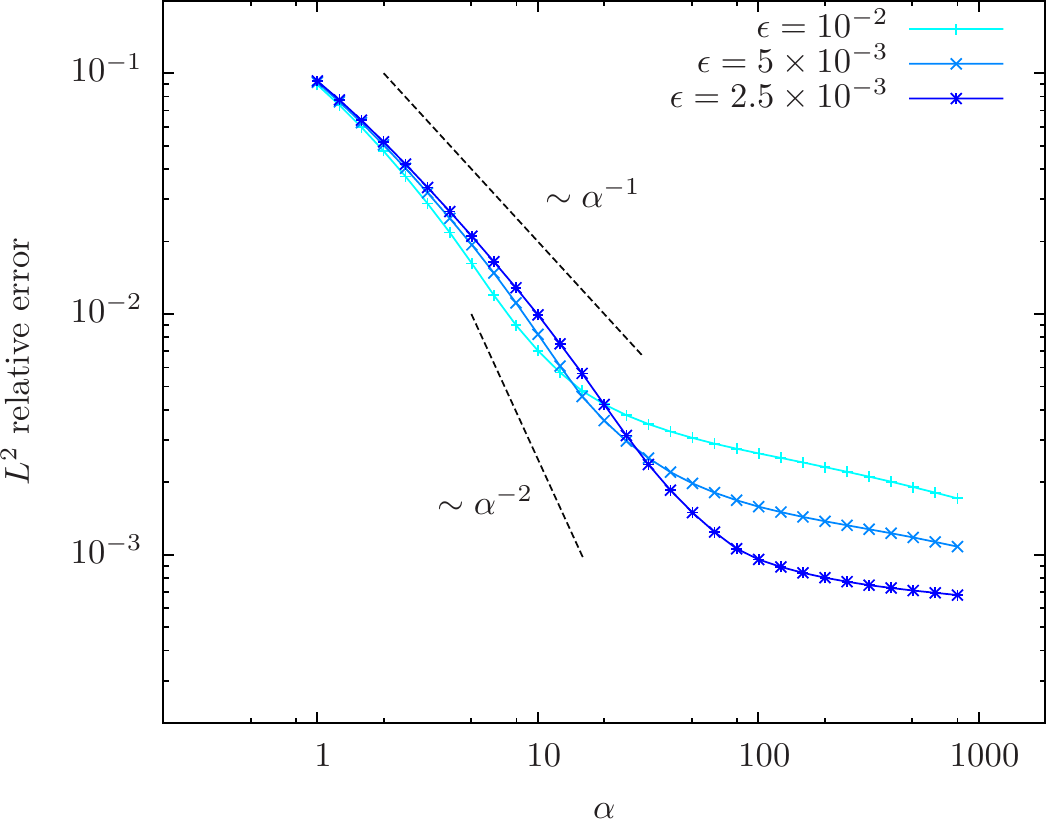}
   \caption{$L^2$ relative error as a function of $\epsilon$ and $\alpha$. The error decreases as $\epsilon\rightarrow0$ and $\alpha\rightarrow\infty$. The vertical dotted lines on the right plot correspond to $St/\alpha=\epsilon$ where the capillary length is equal to the interface thickness. Increasing $\alpha$ further does not improve the convergence of the phase-field model, therefore setting an upper bound for $\alpha$.\label{fig:Aconv}}
\end{figure}

The phase field model introduced in section~\ref{sec:pf} is now used to solve the exact same problem.
Equations~\eqref{eq:pf} and \eqref{eq:temp} are solved on a one-dimensional domain neglecting fluid motions.
The numerical scheme is the same as for the mapping method described above.
The spatial resolution is fixed to $N=256$ for all simulations.
In addition to the physical parameters used above, we need to specify three additional numerical parameters specific to the phase-field formulation: $\epsilon$, $\alpha$ and $m$.
Following \cite{Caginalp1989} and \cite{Wang1993}, the mobility is fixed whereas $\alpha$ and $\epsilon$ are varied. In this paper, all simulations are performed with $m=1$ and we checked that varying $m$ does not qualitatively affect the solution.
The first parameter $\epsilon$ corresponds to the effective thickness of the interface and the original discontinuous problem is obtained taking $\epsilon\rightarrow0$.
In practice, the minimal value of $\epsilon$ is directly related to the grid-size.
We therefore perform several simulations varying $\epsilon$ while all other parameters remain fixed, $m=1$ and $\alpha=200$.
For each value of the interface thickness $\epsilon$, we compute the $L^2$ relative error on the temperature profile at $t=1/4$ between the phase-field model and the reference solution obtained with the mapping method.
Results are shown in Figure~\ref{fig:Aconv}.
The relative error increases with $\epsilon$ with a power law between $\epsilon$ and $\epsilon^2$.
The minimum of the error is obtained for $\epsilon\approx2\times10^{-3}$ which is half of the grid size $\textrm{d}x\approx4\times10^{-3}$.
This is not surprising since the effective thickness of the interface at equilibrium is given by $2\sqrt{2}\epsilon$ (see equation~\eqref{eq:eqp}).
When $\epsilon$ decreases below this critical value of $\epsilon=\textrm{d}x/2$, the phase field model becomes unstable and leads to nonphysical results.

In order to show that the particular choice of the functions $p(\phi)$ and $g(\phi)$ in equation~\eqref{eq:dimpf} is arbitrary and does not affect the results, we also show in Figure~\ref{fig:Aconv} the convergence of the so-called Model II of Wang which only differs from Model I by the choice of $p(\phi)$. Instead of equations~\ref{eq:pdp}, Model II chooses $p(\phi)=\phi^2\left(3-2\phi\right)$. Results are very similar for both models. We nevertheless stick with Model I since this particular choice of functions ensures that all thermodynamic constraints are satisfied independently of the temperature distribution.

We now consider the convergence of the phase-field model as a function of the coupling parameter $\alpha$.
We repeat the same procedure but we now fix $\epsilon=2.5\times10^{-3}$.
$\alpha$ is varied from $1$ to $\alpha=800$.
The same relative error is shown in Figure~\ref{fig:Aconv}.
When $\alpha$ is of order unity, the Stefan boundary condition is not satisfied since the capillary length is too large and the temperature at the interface is not exactly the melting temperature. As $\alpha$ increases, the relative error drops by several order of magnitude, following a power law between $\alpha^{-1}$ and $\alpha^{-2}$.
For very large values of $\alpha$, typically $\alpha>0.2/\epsilon$ the error saturates, which is expected since the capillary length becomes as small as the interface thickness.
This is confirmed by repeating the same experiment for different values of the interface thickness.
Similar results are obtained using the alternative Model II.

In conclusion, this one-dimensional study shows that all numerical parameters introduced by the phase-field model are actually constrained in order to accurately represent the original Stefan problem.
The interface mobility is fixed to an arbitrary value of order unity and we choose here the simplest choice $m=1$.
The interface thickness is taken as small as possible and is related to the grid size $\textrm{d}x$ by $\epsilon=\textrm{d}x/2$.
Once $\epsilon$ and $m$ are fixed, we choose $\alpha=St/\epsilon$ where $St$ is the Stefan number (see equation~\eqref{eq:stefan_bc}).
This approach ensures that for a given spatial resolution, the error with respect to the original Stefan problem is minimized.

\subsection{One-dimensional axisymmetric case without fluid motions\label{sec:A2}}

We now consider an axisymmetric case without motion in the fluid phase.
This additional validation is important since the previous 1D study neglected curvature effects.
The phase-field method introduces an effective surface tension at the interface, which has to be negligible in order to recover the classical Stefan problem.
The boundary condition associated with the phase field model in the limit of vanishing interface thickness is
given by equation~\eqref{eq:stefan_bc} where curvature at the interface and kinetics can modify the temperature at the interface.
In order to recover the Stefan boundary condition, they need to be negligible leading to the limit $\alpha/St\rightarrow\infty$.
Since the forcing term in the phase-field equation~\eqref{eq:pf} is solved explicitly here, $\alpha$ is necessarily bounded for stability reasons.
This means that, for a given problem, there is a maximum curvature of the interface above which the Stefan boundary condition will not be satisfied.
In this paper, we typically take $\alpha/St\sim\epsilon$, so that the upper bound for the curvature is approximately the inverse of the interface thickness.

To study the convergence of the phase-field model as curvature effects become important, we study a simple axisymmetric Stefan problem.
A solid disk of initial radius $r_i$ is immersed inside an infinite fluid domain.
The initial temperature distribution is given by
$\theta(r)=\left[1+\tanh(\gamma(r-r_i))\right]/2$ where $r=0$ is the center of the disk and we choose $r_i=1/10$ and $\gamma=100$.
The solid is therefore mostly at $\theta=0$ while the liquid is mostly at $\theta=1$, the melting temperature being fixed to $\theta_M=1/2$.
The Stefan number is fixed to $St=1$.

\begin{figure}
   \vspace{5mm}
   \centering
   \includegraphics[width=0.47\textwidth]{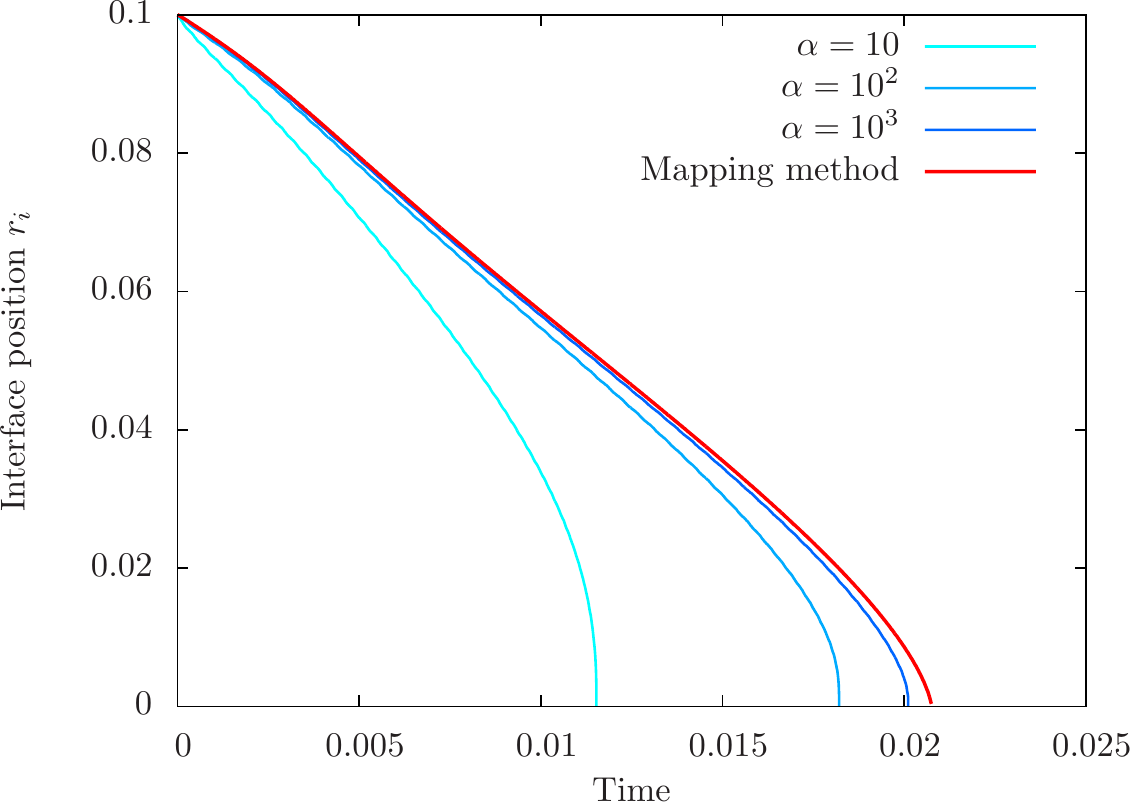}
   \hspace{2mm}
   \includegraphics[width=0.49\textwidth]{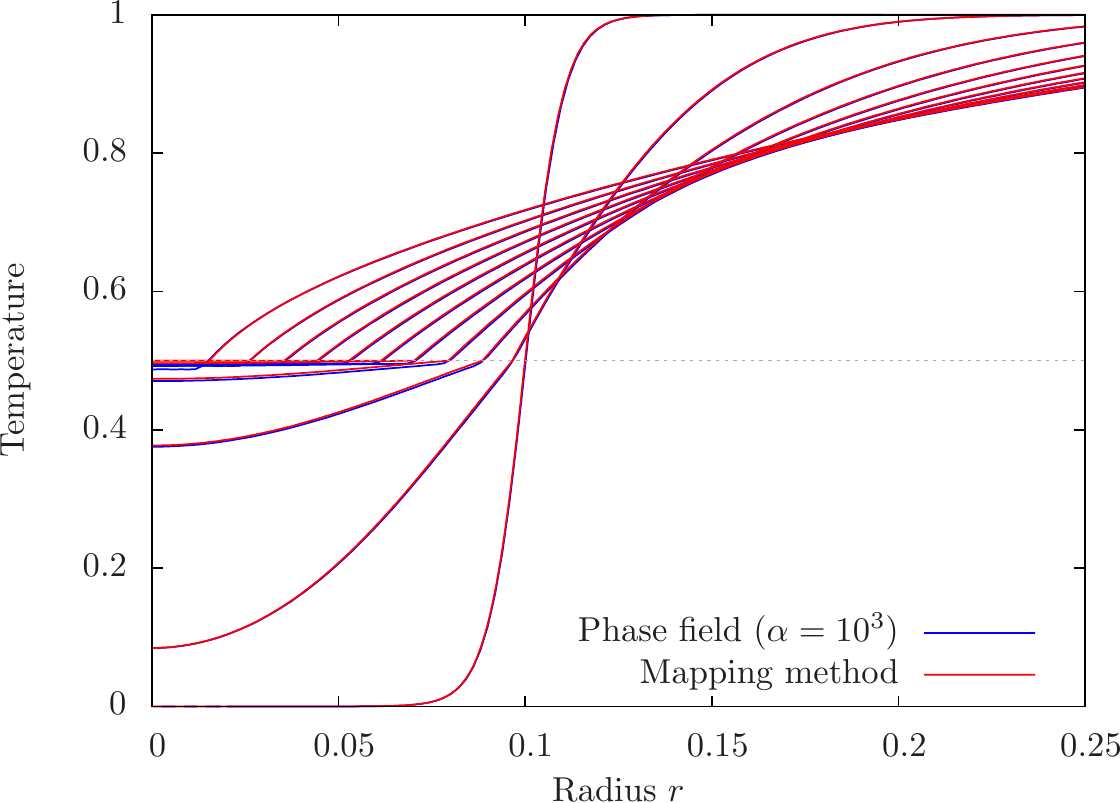}
   \caption{Left: temporal evolution of the interface curvature as obtained by the phase field method with varying $\alpha$. The dots corresponds to the time at which the solid is completely melted. Right: radial profile of temperature in the solid and liquid domains at different times.\label{fig:Acyl}}
\end{figure}

The problem is first solved using the same mapping method as in section~\ref{sec:A1}, solving for the radial diffusion of heat in both liquid and solid domain and explicitly matching the two solutions in order to satisfy the Stefan condition at the interface.
We use $N=64$ grid points for the small solid domain and $N=512$ grid points for the larger liquid domain.
To test the phase-field model, we use the same numerical code as before, adapted to cylindrical coordinates.
We use $N=512$, $\epsilon=10^{-3}$, $m=1$ as usual, and vary $\alpha$ from $\alpha=10$ to $\alpha=10^3$ which corresponds to the limiting case where the capillary length is equal to the thickness of the interface.
We track the interface position which corresponds to $\phi=1/2$.

Figure~\ref{fig:Acyl} shows the time evolution of the interface position, for both the mapping method and the phase field model with different values of $\alpha$.
As time evolves, the curvature of the interface increases continuously so that, for a fixed $\alpha$, the effective boundary condition of the phase field model \eqref{eq:stefan_bc} departs from the expected Stefan boundary condition $\theta(r_i)=\theta_M$.
Note in addition, the velocity of the interface diverges as $r_i\rightarrow0$, so that the second kinetic term in equation~\eqref{eq:stefan_bc} also starts to contribute.

In conclusion, providing that $\epsilon\rightarrow0$ and $\alpha\rightarrow\infty$, the Stefan boundary condition is recovered except in regions with large curvatures, typically of order $1/\epsilon$.
Even for the extreme case of a vanishing solid considered here, the phase field model predicts the time of complete melting within few percents, and most of the errors is due the final period where both curvature and interface velocity diverge.

\begin{figure}
\unitlength \textwidth
\includegraphics[width=0.43\textwidth]{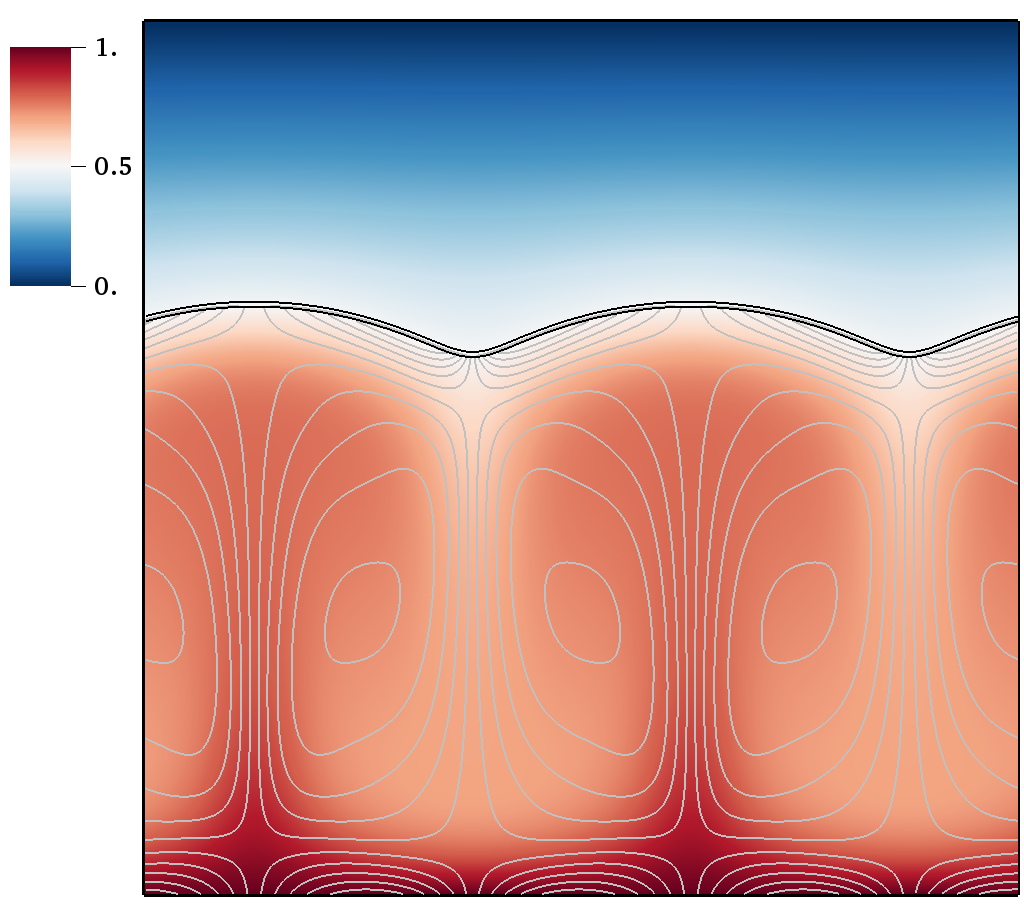}
\hfill
\includegraphics[width=0.52\textwidth]{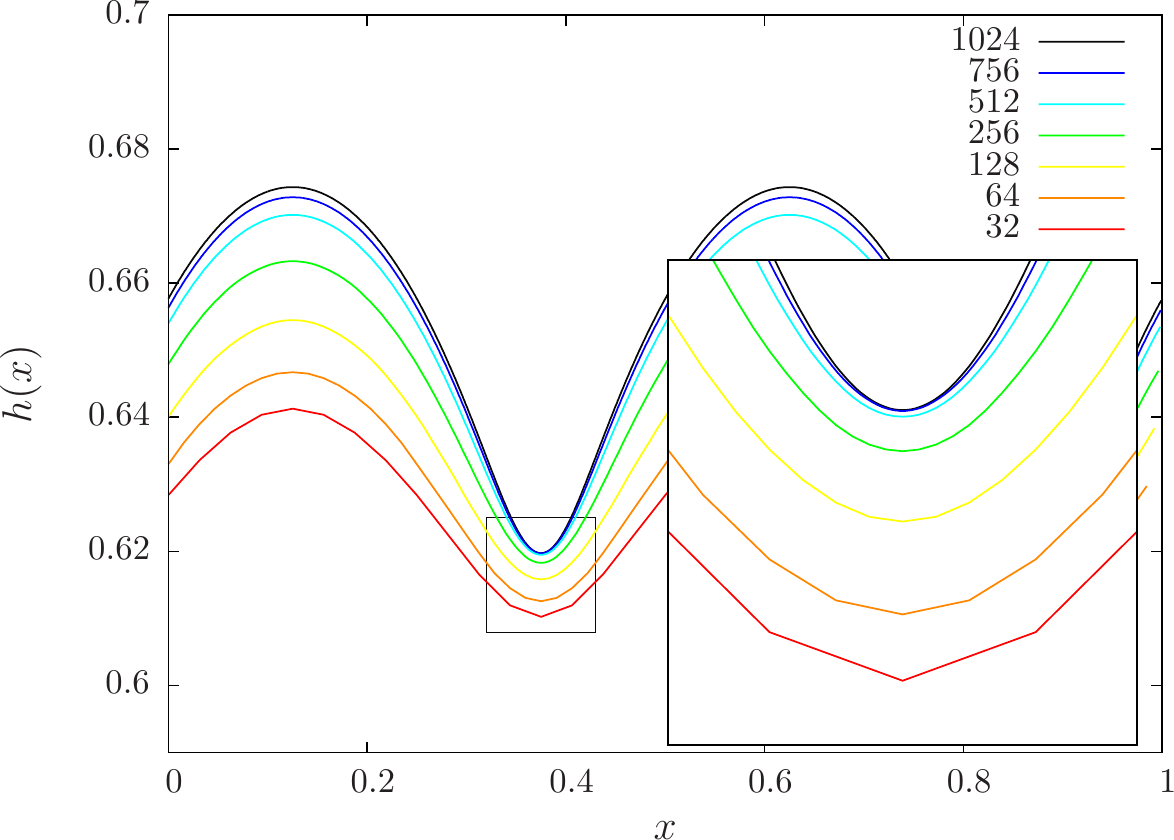}
\caption[]{Left: Visualisation of the temperature field, vorticity and phase field contours for the case $N=1024$ at $t=0.1$. Right: interface position $h(x)$ as a function of the resolution $N$ and at $t=0.1$.}
\label{fig:A7}
\end{figure}

\subsection{Two-dimensional case with fluid motions\label{sec:a3}}

Finally, let us discuss the complete problem where fluid motion occurs in the fluid phase.
To our knowledge, there is no analytical solution of such a problem, so that we perform a relative convergence study, where our reference is a numerical simulation with the largest available resolution.
We consider a two-dimensional square domain of unit side, with periodic boundary conditions in the $x$-direction and no-slip fixed temperature boundary conditions in the $z$-direction.
We choose the simple initial equilibrium given by $\theta(z)=1-z$ where $\theta_M=1/2$, the interface being initially located at $z=1/2$.
To this equilibrium state, we add a temperature perturbation inside the liquid domain of the form $\theta'(x,z)=A\sin(4\pi x)\sin^2(2\pi z)$ and $\theta'(x,z)=0$ in the solid domain.
We choose the following set of physical parameters: $Ra=5\times10^5$, $St=1$, $Pr=1$.

This problem is solved using an increasing spatial resolution from $N=32$ to $N=1024$ in each spatial direction.
The highest resolution is our reference case to which we compare all other resolutions.
The phase-field and penalization parameters are automatically adjusted depending on the resolution according to the following rules.
The interface thickness is chosen as small as possible and is directly proportional to the grid-size $\epsilon=1/(N-1)$ (see appendix~\ref{sec:A1}).
The coupling parameters $\alpha$ is taken to be $St/\epsilon$ and the mobility $m=1$.
The penalization parameter is automatically chosen to be $\eta=2\textrm{d}t$ where $\textrm{d}t$ is the time-step.
We let the simulation evolves up to $t=0.1$.

\begin{figure}
   \vspace{5mm}
   \centering
   \includegraphics[width=0.7\textwidth]{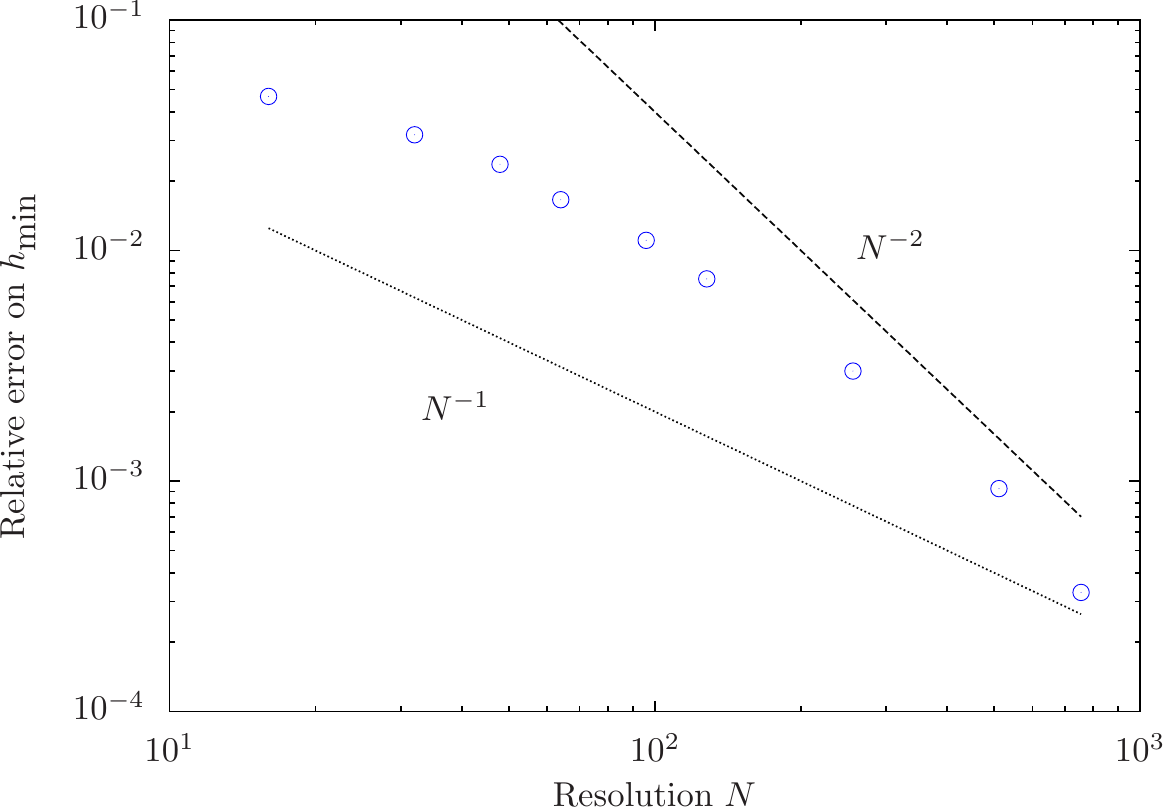}
   \caption{Relative error on the minimum value of the interface position at $t=0.1$ as a function of the spatial resolution. The reference case corresponds to the case $N=1024$.\label{fig:A8}}
\end{figure}

A snapshot of the reference solution with $N=1024$ is shown in Figure~\ref{fig:A7}.
The colors correspond to the temperature field, the grey lines correspond to contours of the vorticity and the two black lines correspond to contours of the phase field characterized by $\phi=0.1$ and $\phi=0.9$.
The interface position $h(x,t)$ is computed at each iteration and for each horizontal position by interpolating the temperature field in order to find the vertical position of the isotherm $\theta=\theta_M=1/2$.
This is achieved using a fourth-order Lagrangian interpolation scheme in the vertical direction only.
The position of the interface at $t=0.1$ for all resolutions considered here is shown in the right panel of Figure~\ref{fig:A7}.
The characteristic topography is observed for all resolution, and its vertical position is converging toward an asymptotic solution as $N$ is increased.
The insert figure shows details of the topography close to a cusp which also corresponds to the slowest convergence of our scheme due to high curvature effects (see appendix~\ref{sec:A2}).
Nevertheless, we compute the relative error on the lowest value of $h(x)$ between the reference case at $N=1024$ and all other cases.
Results are shown in Figure~\ref{fig:A8}, where a second order convergence with the resolution is obtained.
Note that the slow convergence, typically first order, obtained at low resolution is due to the overlap between the typical size of the thermal boundary layers and the typical size of the interface $\epsilon$ (which is, we recall, enslaved to the spatial resolution here).
In other words, this confirms the intuitive conclusion that the interface thickness must be smaller than any other physical length scales for the phase field model to be appropriate.
All simulations presented in this paper satisfy such a condition.

\section{Second-order correction of the diffusion flux with topography\label{sec:appC}}

Using expression \eqref{h_exp}, the expanded boundary condition \eqref{upper_bound} reads:
\begin{eqnarray}
\theta_M & = & \theta(x,h_0) + \varepsilon h_0 \cos{kx} \frac{\partial \theta}{\partial z}(x,h_0) 
+ 
\varepsilon^2 h_0^2 \cos^2{kx} \frac{\partial^2 \theta}{\partial z^2}(x,h_0)
+
\cdots\\ 
& = & \theta_0(x,h_0) \nonumber\\
& + & \varepsilon 
\left( 
\theta_1(x,h_0) + 
h_0 \cos{kx} \frac{\partial \theta_0}{\partial z}(x,h_0) 
\right) \nonumber\\
& + &
\varepsilon^2 
\left( 
\theta_2(x,h_0) + 
h_0 \cos{kx} \frac{\partial \theta_1}{\partial z}(x,h_0)
+
h_0^2 \cos^2{kx} \frac{\partial^2 \theta_0}{\partial z^2}(x,h_0)
\right)
+
\cdots
\end{eqnarray}

The leading-order solution of the diffusion equation for $\varepsilon=0$ is simply given by:
$$
\theta_0(x,z)= 1 - \Delta \theta \frac{z}{h_0} \ .
$$
At order $\varepsilon$, we need to solve:
$$
\nabla^2 \theta_1 = 0 \ ,
$$
with the boundary conditions:
\begin{eqnarray}
\theta_1(x,0) & = & 0\\
\theta_1(x,h_0) + h_0 \cos{kx} \frac{\partial \theta_0}{\partial z}(x,h_0) & = & 0 \ . \label{upper_bound2}
\end{eqnarray}
The solution reads:
$$
\theta_1(x,z) = \Delta \theta \cos{kx} \frac{\sinh{ky}}{\sinh{kh_0}} \ .
$$
At order $\varepsilon^2$, we need to solve:
$$
\nabla^2 \theta_2 = 0 \ ,
$$
with the boundary conditions:
\begin{eqnarray}
\theta_2(x,0) & = & 0\\
\theta_2(x,h_0) + 
h_0 \cos{kx} \frac{\partial \theta_1}{\partial z}(x,h_0)
+
h_0^2 \cos^2{kx} \frac{\partial^2 \theta_0}{\partial z^2}(x,h_0)
& = & 0 \ . \label{upper_bound3}
\end{eqnarray}

The solution reads:
$$
\theta_2(x,z) = -\frac{k \Delta \theta}{2} \coth{k h_0} 
\left(
z + h_0 \cos{2kx} \frac{\sinh{2kz}}{\sinh{2kh_0}}
\right) \ ,
$$
and the full second-order solution is:
$$
\theta(x,z) = \theta_0(x,z) + \varepsilon \theta_1(x,z) + \varepsilon^2 \theta_2(x,z) \ .
$$
Finally, the mean heat flux through the lower boundary can be computed from this expansion. 
The only two contributions to the flux come from $\theta_0$ and $\theta_2$:
\begin{eqnarray}
Q_D 
& = &
\frac{\Delta \theta}{h_0}
+\varepsilon^2 \frac{k \Delta \theta}{2} \coth{k h_0}
\frac{1}{\lambda} \int_0^\lambda {
\left(
1 + 2 k h_0 \cos{2kx} \frac{1}{\sinh{2kh_0}}
\right)
}dx \nonumber\\
& = & 
\frac{\Delta \theta}{h_0}
+\varepsilon^2 \frac{k \Delta \theta}{2} \coth{k h_0}
\label{eq:th_hf}
\end{eqnarray}

\begin{figure}
   \vspace{5mm}
   \centering
   \includegraphics[width=0.7\textwidth]{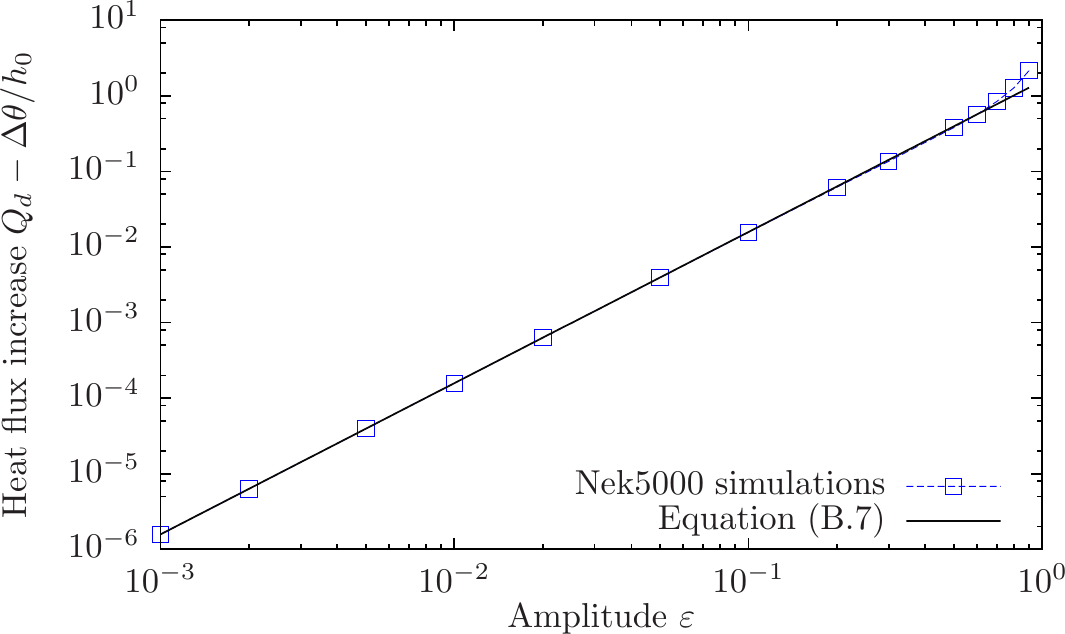}
   \caption{Comparison between the prediction from equation~\eqref{eq:th_hf} and direct numerical simulations using the spectral element solver Nek5000.\label{fig:hi}}
\end{figure}

We compare this scaling with actual numerical simulations of the heat flux through a plane layer with the upper topography being defined by equation~\eqref{h_exp}.
These simulations were performed using the spectral element solver Nek5000 \citep{Nek5000} which can easily accommodate this kind of non-trivial geometries.
Nek5000 has for example been recently used to study flows inside tri-axial ellipsoids \citep{Grannan2017}.
We use a 2D Cartesian mesh made of $256$ elements and a polynomial order of $12$.
The aspect ratio of the box is $\lambda=2$ and we explicitly impose the topography given by equation~\eqref{h_exp}.
The horizontal boundaries are periodic and we fix the temperature on both the planar bottom and corrugated upper boundaries.
We run several simulations varying the amplitude $\varepsilon$ and we measure the heat flux \eqref{eq:thf} in the steady state for $Ra=0$ to ensure a purely diffusive regime.
The agreement between the theoretical prediction \eqref{eq:th_hf} and the direct measure of the diffusive flux $Q_D$ in the Nek5000 simulations is excellent as can be seen in Figure~\ref{fig:hi}, up to topographies with amplitude around $\varepsilon\approx0.5$, well below the typical amplitudes observed in our melting simulations, thus justifying the use of the reference diffusive heat flux~\eqref{eq:th_hf} to defined the Nusselt number.
Note that if instead of imposing a simple monochromatic topography, we consider a more realistic configuration involving cusps for example, the correction to heat flux remains of second order, and only the prefactor is slightly modified.

\bibliography{biblio}
\bibliographystyle{jfm}

\end{document}